\documentclass[iop]{emulateapj}

\newcommand{\beq}{\begin{equation}}
\newcommand{\eeq}{\end{equation}}

\newcommand{\teff}{$T_{\rm eff}$}
\newcommand{\lsim}{\ \raise
-2.truept\hbox{\rlap{\hbox{$\sim$}}\raise5.truept\hbox{$<$}\ }}
\newcommand{\gsim}{\ \raise
-2.truept\hbox{\rlap{\hbox{$\sim$}}\raise5.truept\hbox{$>$}\ }}
\newcommand{\simsim}{\ \raise
-2.truept\hbox{\rlap{\hbox{$\sim$}}\raise5.truept\hbox{$\sim$}\ }}

\def\arcmin{\hbox{$^\prime$}}

\def\deg{\hbox{$^\circ$}}
\def\MH{\hbox{\sl [M/H]}}

\usepackage{rotating}

\slugcomment{Accepted for publication in The Astrophysical Journal Supplement Series}
\shorttitle{Optical Photometry of the ONC. II}
\shortauthors{Da Rio et al.}

\begin{document}

\title{A Multi-color Optical Survey of the Orion Nebula Cluster. II. the H-R diagram}

\author{N. Da Rio\altaffilmark{1}}
\affil{Max Planck Institut f\"{u}r Astronomie,  K\"{o}nigstuhl 17, D-69117 Heidelberg, Germany }
\altaffiltext{1}{Member of IMPRS for Astronomy \& Cosmic Physics
       at the University of Heidelberg}
\email{dario@mpia-hd.mpg.de}

\author{M. Robberto, D. R. Soderblom, N. Panagia\altaffilmark{2,3}}
\affil{Space Telescope Science Institute, 3700 San Martin Dr., Baltimore MD, 21218, USA\ }
\altaffiltext{2}{INAF-CT Osservatorio Astrofisico di Catania, Via S.Sofia 79, I-95123 Catania, Italy}
\altaffiltext{3}{Supernova Ltd, OYV \#131, Northsound Road, Virgin Gorda, British Virgin Islands}

\author{L. A. Hillenbrand}
\affil{California Institute of Technology, 1200 East California Boulervard, 91125 Pasadena, CA, USA}

\author{F. Palla}
\affil{INAF - Osservatorio Astrofisico di Arcetri, Largo E. Fermi, 5 I-50125 Firenze, Italy}

\author{K. G. Stassun\altaffilmark{4}}
\affil{Vanderbilt Univ., Dept. of Physics \& Astronomy 6301 Stevenson Center Ln., Nashville, TN 37235, USA\ }
\altaffiltext{4}{Fisk University, Department of Physics, 1000 17th Ave. N., Nashville, TN 37208, USA}


\begin{abstract}
We present a new analysis of the stellar population of the Orion Nebula Cluster (ONC) based on multi-band optical photometry and spectroscopy.
We study the color-color diagrams in $BVI$, plus a narrow-band filter centered at 6200\AA, finding evidences that intrinsic color scales valid for main-sequence dwarfs are incompatible with the ONC in the M spectral type range, while a better agreement is found employing intrinsic colors derived from synthetic photometry, constraining the surface gravity value as predicted by a pre-main sequence isochrone. We refine these model colors even further, empirically, by comparison with a selected sample of ONC stars with no accretion and no extinction. We consider the stars with known spectral types from the literature, and extend this sample with the addition of 65 newly classified stars from slit spectroscopy and 182 M-type from narrow-band photometry; in this way we isolate a sample of about$~1000$ stars with known spectral type. We introduce a new method to self-consistently derive the stellar reddening and the optical excess due to accretion from the location of each star in the $BVI$ color-color diagram. This enables us to accurately determine the extinction of the ONC members, together with an estimate of their accretion luminosities. We adopt a lower distance for the Orion Nebula than previously assumed, based on recent parallax measurements. With a careful choice also of the spectral type-temperature transformation, we produce the new Hertzsprung-Russell diagram of the ONC\ population, more populated than previous works. With respect to previous works, we find higher luminosity for late-type stars and a slightly lower luminosity for early types. We determine the age distribution of the population, peaking from $\sim2$ to $\sim3$~Myr depending on the model, a higher age than previously estimated. We study the distribution of the members in the mass-age plane, and find that taking into account selection effects due to incompleteness removes an apparent correlation between mass and age. We derive the IMF for low- and intermediate-mass members of the ONC, which turns out to be model-dependent, and shows a turn-over at $ M\lesssim 0.2~$${\rm M_\odot}$.
\end{abstract}


\section{Introduction}
\label{section:introduction}
The Orion Nebula (M42, NGC~1976), with its associated young stellar cluster (Orion Nebula Cluster, ONC),\ is the nearest site of recent massive star formation and is therefore  regarded as a prototype of dense star forming regions. At a distance of $\sim 400$~pc from the Sun \citep{sandstrom07,menten2007,hirota07}, the region is projected on the northwestern side of the Orion A giant molecular cloud, but contrary to the majority of star forming regions in the Galaxy   \citep{ladalada2003}, it is only slightly extinguished. This is due to the presence of the most massive members of the young stellar population,  $\theta^1$Ori-C in the Trapezium cluster in particular, whose ultra-violet emission has carved an ionized cavity in the molecular material which now engulfs most of the young cluster. The high  extinction provided by the background molecular cloud (up to $A_V\simeq100$~mag, \cite{bergin96}),  the position of this region at the anticenter of the Galaxy ($l \simeq 209\deg$), and the high projected distance from the galactic plane ($b \simeq-19\deg$) conspire to provide a rather low foreground and background stellar contamination, enabling us to perform extinction-limited studies of a young, nearly coeval,  relatively rich and uncontaminated pre-main sequence (PMS) cluster ($1-3\times 10^{4}$ stars aged $\sim1-3$~Myr, \citealp{hillenbrand2001}).

Despite these advantages, observational studies of the ONC are not without  difficulties. The nebulosity which surrounds the ONC population and the presence of circumstellar material (disks and envelopes) cause highly non-uniform extinction among the members which must be corrected for each individual star. Furthermore, on-going accretion, variability due to star spots or stellar instabilities, influence the observed optical colors, which may vary with time. In principle, an observational effort including simultaneous multi-band photometry over a large wavelength range and adequate temporal sampling to account for variability effects may overcome these difficulties, disentangling the non-photospheric emission from complete spectral energy distributions (SEDs). Even in the presence of perfect measurements of photometry, and with consideration of the above effects, uncertainties remain in the conversions between colors and magnitudes to temperatures and luminosities, which differ from those of main sequence stars especially for late spectral types.

The combination of these uncertainties make the construction of an accurate H-R diagram (HRD) based on photometry alone for the PMS population misleading. The construction of the HRD is more robust when photometry is complemented with spectroscopy, since spectral types can be easily determined even though moderate extinction, and the accretion excess can be directly identified from the line emissions or measurement at high dispersion of spectral veiling. Nonetheless, also in this case, the uncertainty in the spectral type-\teff\ relation and, in addition, the difficulty of obtain spectra in a crowded field with bright, variable background emission limit the accuracy in deriving the stellar parameters.
The ONC has been repeatedly studied both at visible wavelengths \citep[e.g.,][]{herbig86,1994ApJ...421..517P,hillenbrand97,robberto04} and in the near-IR \citep[e.g][]{ali-depoy95,hillenbrand98disks,hillenbrand2000,lucasroche2000,luhman2000,slesnick04,robberto09} with increasing field coverage and sensitivity, probing the cluster well into the substellar regime. From the point of view of constructing the HRD of the cluster, the most complete work remains that of \citet{hillenbrand97} (hereafter H97). Using compiled plus new $VI$ photometry and optical spectroscopy, H97 derived the extinction for $\sim 1000$ stellar (i.e., above the H-burning mass)\ members and used the dereddened photometry to place the stars in the HRD. This made it possible to derive the stellar ages and masses through the comparison with PMS evolutionary models.

In a previous paper \citep[][hereafter Paper~I]{paperI} we have presented the optical photometry of the ONC obtained in $UBVI$, H$\alpha$ and in the $6200\AA$ medium-band filters using the Wide Field Imager (WFI) at the 2.2m MPG/ESO telescope on La~Silla. In this paper we use these data to construct the HRD for the intermediate- ($1$~M$_{\odot}\lesssim M \lesssim 8$~M$_{\odot}$) and low-mass ($0.08$~M$_{\odot}\lesssim M \lesssim 1$~M$_{\odot}$) end of the stellar population. As in H97, we complement the optical photometry with spectroscopy, extending the spectral catalog of that work with new spectral types from slit spectroscopy, and also from narrow-band photometry. This allows us to improve over H97 extending the completeness of the sample of members placed in the HRD. The use of multi-color photometry provides advantages in estimating the contributions of both accretion luminosity and reddening to the observed colors. However, this is possible only together with an independent estimate of \teff. For late-type stars it is not enough to rely on stellar photometry, in particular on the $V$ and $I$ bands alone, due to the degeneracy between reddening and \teff\ (H97). In principle, multi-band photometric observations can help in breaking the degeneracy \citep{romaniello2002}, but the presence of non-photospheric emission, such as, for example, the $U$ and $B$ excesses due to accretion, limits the accuracy of the results.
The most reliable method to break the temperature-reddening degeneracy remains to obtain an independent estimate of the effective temperature of the stars by means of spectroscopy. Thus we derive the stellar parameters only for the sub-sample with available \teff\ estimates.

The paper is organized in the following way: in Section \ref{section:thephotometry} we summarize the dataset used, which is a $UBVI$ and $TiO$-band photometry from ground based observations; in Section \ref{sectioncolorcolormain} we study the color-color diagrams for all sources in the catalog in comparison with intrinsic colors computed from synthetic photometry; this allows us to test and correct empirically the models in order to derive the intrinsic, photospheric, colors valid for the ONC. In Section \ref{section:subsampleanalysis} we isolate the sub-sample for which a measurement of \teff\ is available, either from optical spectroscopy or narrow-band photometry, and using this subsample, in Section \ref{sectionH-R} we derive the new HRD of the ONC, and discuss the age distribution, the mass function, and the correlation between ages and masses using two families of PMS evolutionary models. In section~\ref{section:conclusion}\ we summarize our findings.

\section{The photometry}
\label{section:thephotometry}
We utilize our multiband optical catalog presented in Paper~I to construct the HRD of the ONC. The catalog  consists of a nearly-simultaneous photometry over a field of view of $\sim34\arcmin\times34\arcmin$ which comprises most of the ONC. We detect $2612$ point-like sources in the $I$ band; 58\%, 43\%, and 17\% of them are also detected in $V$, $B$, and $U$, respectively. $1040$ sources are identified in the H$\alpha$ band, and in the TiO narrow-band at $6200$\AA.

The photometry is not color-corrected towards a standard (e.g, Johnson-Cousins) system, in order to avoid an additional uncertainty introduced by these transformations, which would be prominent and non-linear due to the large difference of the WFI filter bandpasses compared to the standard bands. Therefore our $UBVI$ catalog follows the WFI instrumental photometric system, calibrated only for zero-point derived from comparison to standard field observations in order to conform to the VegaMag convention\footnote{i.e. according to our photometric calibration, Vega has zero apparent magnitude in all the WFI bands.}.

\begin{deluxetable}{crrrr}
\tablecaption{Effective wavelength, FWHM, E.W. and Vega zero-points for the WFI photometric bands used} 
\tablehead{\colhead{Band} & \colhead{$\lambda_{\rm eff}$} & \colhead{FWHM} & \colhead{E.W.} & \colhead{Vega Flux} \\
\colhead{} & \colhead{(\AA)} & \colhead{(\AA)}  & \colhead{(\AA)} & \colhead{erg s$^{-1}$cm$^{-2}$\AA$^{-1}$} }
\startdata
$U$ & 3470 & 668  & 639 & $3.40\cdot10^{-9}$\\
$B$ & 4541 & 1139 & 1058  & $5.88\cdot10^{-9}$\\
$V$ & 5375 & 884  & 791 & $3.86\cdot10^{-9}$\\
$I$ & 8620 & 1354 & 1431 & $9.50\cdot10^{-10}$\\
TiO & 6217 & 193  & 190 & $2.47\cdot10^{-9}$
\enddata
\label{table:filters}
\end{deluxetable}
In Table \ref{table:filters} we summarize the characteristics of the WFI bands we use. For each band the effective throughput is calculated multiplying the filter passbands\footnote{http://www.eso.org/lasilla/instruments/wfi/inst/filters/} by the average CCD quantum efficiency\footnote{http://www.eso.org/lasilla/instruments/wfi/inst/filters/WFI-Filters-2006-09-18.tar.gz} (Q.E.). This is analogous to Table~1 in Paper~I, which included, by way of illustration, central wavelength and FWHM of the filters, but not accounting for the Q.E. The averaged flux of Vega for each band \beq\int{f_{\rm Vega}S_\lambda{\rm d}\lambda}/\int{S_\lambda{\rm d}\lambda}\eeq is presented in column 5 of Table \ref{table:filters}, where $S_\lambda$ is the band throughput and $f_{\rm Vega}$ a calibrated spectrum of Vega \citep{bohlinvega}. This quantity, multiplied by the equivalent width (column 3), provides the zero-points of the photometric systems. For each band, this is the flux in erg s$^{-1}$ cm$^{-2}$ from a zero magnitude source when the filter throughput is normalized to have 100\% response at the peak.

As described in Paper~I, the 50\% completeness limit of our photometric catalog is at $V\simeq 20.8$. In the same work we have also presented a spectral classification of 217 M-type stars, based on a photometric index derived from the $V$, $I$, and TiO magnitude.

The main improvements upon the photometric catalog of H97, which shares a similar field of view of our optical photometry, are a slightly deeper detection limit, a smaller pixel size, a higher effective wavelength of the WFI $I$-band than the $I_c$ band, closer to the peak of the SED for the latest spectral types. Moreover, whereas the H97 photometry was obtained merging data from different instruments and epochs, our catalog is fully based on WFI observations carried out in a single night. In this way the risk of introducing an additional uncertainty from different systematic effects affecting parts of the stellar sample is removed.

\section{Analysis of the photometric colors of the ONC}
\label{sectioncolorcolormain}
For a star of a given spectral type, the broad-band colors as well as the integrated flux and bolometric correction (BC), depend in general on the \teff, $\log g$ and [Fe/H]. For main sequence and post-main sequence stars the dependence has been investigated both through empirical studies \citep{flower96,alonso99}, relying on stellar templates, and through stellar atmosphere modeling (e.g., \citet{allard95}).  For young PMS stars, unfortunately, an extensive empirical study of the intrinsic photospheric colors, and their possible variation compared to MS dwarfs, has never been carried out. This is mostly due to the fact that the observed fluxes for this class of stars are strongly contaminated by excesses caused by accretion activity and circumstellar material; moreover young (few Myr old) PMS populations are usually somewhat embedded into their parental cloud, with extinction differentially affecting the cluster members. As a consequence it is rather difficult to accurately measure the intrinsic photospheric emission of PMS stars.

Further, the effects of non-uniform foreground extinction, those from the amount and composition of any circumstellar dust which affects both the total extinction and the reddening law, and those from the possible presence of on-going mass accretion cause a combination of reddening and blueing effects beyond the uncertainties in the intrinsic colors.   The dependence of observed colors on the various physical parameters involved is degenerate.
For these reasons, optical photometry alone is insufficient to derive stellar parameters for young stars
and additional constraints are needed, for instance an independent estimate of \teff\ for the star.
In order to disentangle or at least constrain the various effects, a precise knowledge of the
underlying photospheric emission -- i.e., the intrinsic colors -- of young stars is required; we focus our discussion here first.

Our multi-band photometry on the ONC can inform us as to the behavior of the optical photospheric colors of young stars. The observed population spans a broad range in \teff and also features a range of disk and accretion properties, including diskless and non-accreting objects which can be used to infer intrinsic colors as a function of \teff.  In this section we compare intrinsic colors predicted using synthetic photometry (Section \ref{sectionsyntheticcmd}), with the observed colors of stars in the ONC. This allows us to both test the accuracy of the model predictions, and to constrain empirically those valid for the ONC (Section \ref{section:emprical_colors}).  Additionally, we use synthetic photometry to model the effects of extinction and mass accretion on the integrated colors. The results of this analysis will be used in Section \ref{section:subsampleanalysis} to derive the stellar parameters of the members, for the subsample of the catalog with available \teff estimate.

\subsection{Synthetic photometry}
\label{sectionsyntheticcmd}

Given the wide range of stellar spectral types covered by our survey, spanning from early types to late-M stars, we must consider different families of synthetic spectra which cover different ranges of \teff. The selection of adequate model atmospheres is particularly critical for late type stars (M-type and later), whose SEDs are dominated by broad molecular absorption features and possibly contain some dust in their atmospheres that can affect the observed spectra and colors.

\begin{figure}
\epsscale{1.1}
\plotone{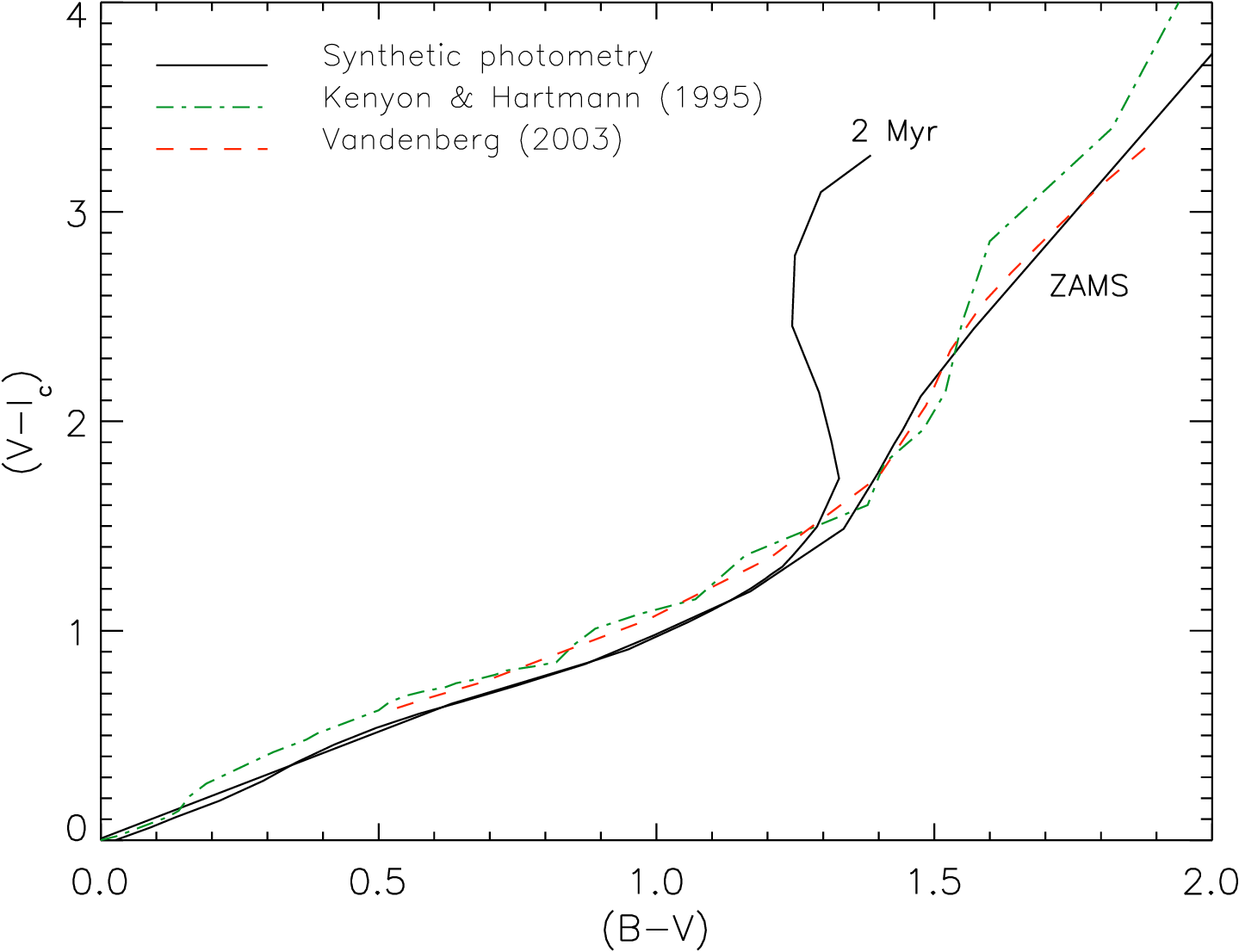}
\caption{Comparison between the intrinsic colors derived by our synthetic photometry for a 2~Myr and the ZAMS (black lines) and the empirical colors valid for MS dwarfs from \citet{kenyon-hartmann95} and \citet{vandenberg2003}, in the Johnson-Cousins $BVI_c$ bands. The modeled ZAMS, compatible with the empirical counterpart, suggests a fairly good accuracy of the atmosphere models we used, at least for $\log g$ values of dwarfs. \label{fig:ZAMS_comparison_empirical_synthetic}}
\end{figure}

For intermediate temperatures we used the NextGen models \citep{nextgen}. This grid of synthetic spectra, computed in the range $3000$ K$<$\teff$<10000$ K using the stellar atmosphere code PHOENIX \citep{hauschildt-phoenix} is known to reproduce fairly well the optical colors of stars down to the K spectral class. For late-M stars these models are not adequate to reproduce the observational data. \citet{BCAH98} show that they are unable to fit solar-metallicity $V$ vs $(V-I)$ color-magnitude diagrams for main sequence populations. The problems set in for $(V-I)\gtrsim2$, which corresponds to \teff$\lesssim3600$ K, and can be attributed to  shortcomings in the treatment of molecules. In particular, \citet{allardtio} illustrate that the main reason for the failure of NextGen spectra in matching the optical properties of late-type stars is the incompleteness of the opacities of TiO lines and water. In that work they present new families of synthetic spectra (AMES-MT models)\ with updated opacities; the optical colors computed from these models were tested against the $VRI$ photometric catalog of \citet{leggett92}, finding a fairly good match with the data down to the lowest temperatures. This suggests some confidence in the ability of the AMES models to reproduce the optical colors of main sequence dwarfs, at least in the $V-I$ range. We anticipate that in this section (see Figure \ref{fig:ZAMS_comparison_empirical_synthetic}) we will further demonstrate the good agreement between the synthetic ZAMS from AMES models and empirical $BVI$ color scales valid for MS dwarfs. Although it remains to be demonstrated that the AMES models are appropriate also for the lower gravities of the PMS stars, the match for the main sequence gives us some confidence in the use of this grid of spectra.

We therefore use the AMES-MT models for the low temperatures in our set of atmosphere models instead of the NextGen spectra; although this substitution is strictly required only for \teff$\lesssim4000$ K, we adopt it for the whole $2000$ K$<T_{\rm eff}<5000$ K range so to have a smooth connection with NextGen at intermediate temperatures. For \teff$>8000$ K we use the Kurucz models \citep{kurucz93}. Due to source saturation, we lack the most luminous members of the ONC in our photometric catalog, and therefore our sample is not populated in the early type end of the population. The metallicity adopted is [M/H]$=0$, the only value available for the AMES-MT grid, which is in good agreement with the observations of ONC members \citep[e.g.][]{dorazi2009}.

\begin{figure}
\epsscale{1.1}
\plotone{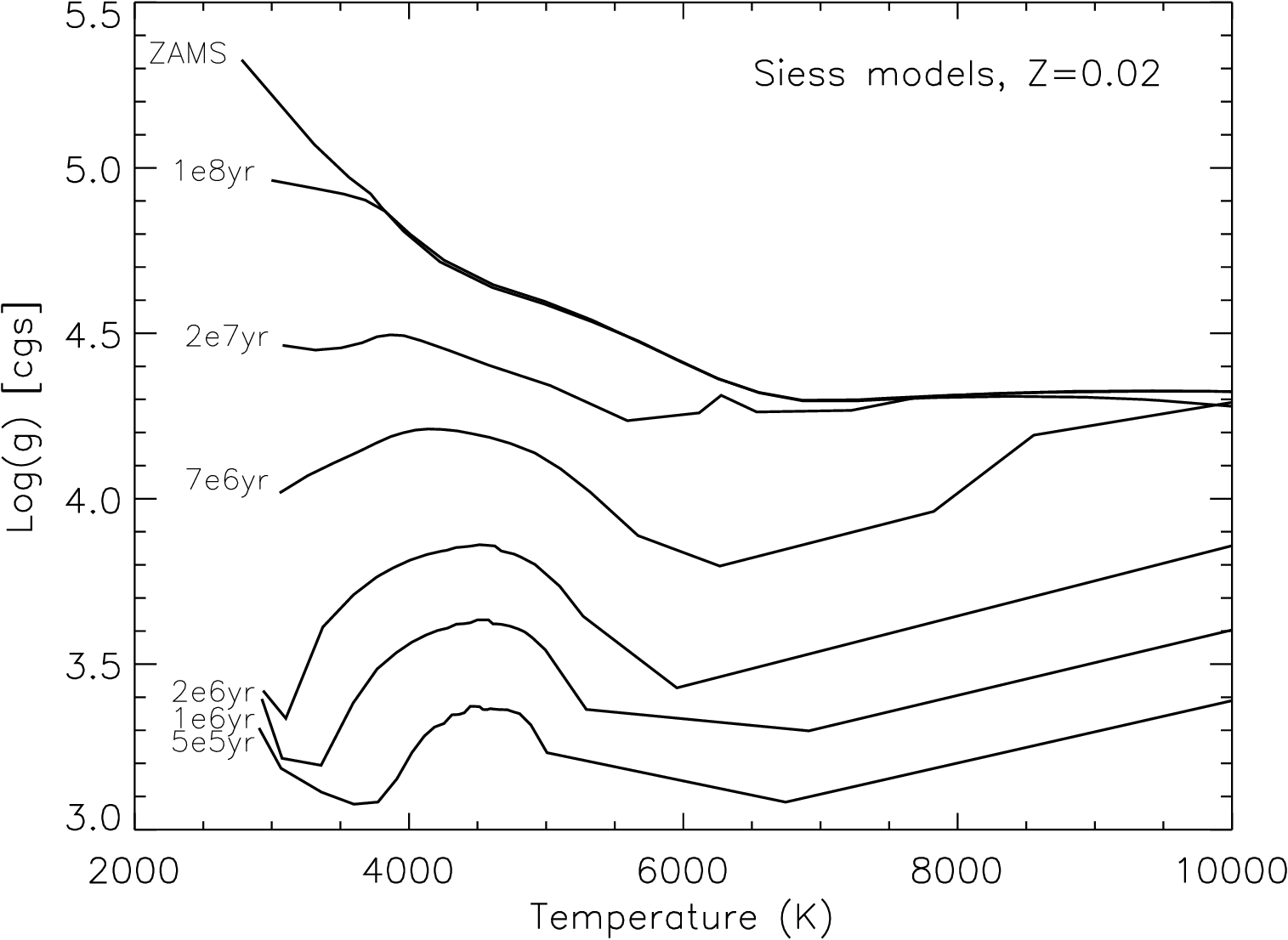}
\caption{$\log g$ vs. \teff\ relations for a sample of Siess isochrones.  \label{fig:isochloggt}}
\end{figure}

The values of $\log g$ span from 3.0 to 5.5 (in logarithmic scale with $g$ in cgs units) with a step of 0.5 dex.
During the PMS phase stellar contraction leads to a decrease of the stellar radius and therefore an increase in $\log g$ according to the relation:
\begin{equation}
\label{eq:logg_t}
\log g = \log\bigg[\frac{M}{{\rm M}_{\odot}}\bigg(\frac{R}{R_{\odot}}\bigg)^{-2}\cdot g_{\odot}\bigg]
\end{equation}
and $g_\odot=2.794\cdot10^4$ g cm s$^{-2}$. Examples of these relations are shown in Figure \ref{fig:isochloggt} for \citet{siess2000} isochrones from $0.5$ Myr to the ZAMS.

In general increasing ages correspond to increasing surface gravities and for the M-type stars (\teff$\lesssim$4000 K) the difference in $\log g$ can be more than two orders of magnitude during the PMS evolution, changing from $\log g\simeq3.5$ for a star aged a few Myr to $\log g\simeq5.5$ for a dwarf M-type stars. For early spectral types the difference is lower, about 1~dex ($\log g\simeq 3.5$ to $\log g\simeq 4.5$). We therefore create, by interpolating within our grid of synthetic spectra, a family of spectra corresponding to each Siess isochrone, as a function of temperature. On this family we perform synthetic photometry to compute absolute magnitudes in the WFI instrumental photometric system, the standard of the photometric catalog presented in Paper I.

The absolute magnitude in a photometric band $S_{\lambda}$ of a star with a spectral energy distribution $F_{\lambda}$, and stellar radius $R$ is given by
\begin{equation} \label{equation:first}
M_{S_{\lambda}}=-2.5\log\bigg[\bigg(\frac{R}{10\textrm{pc}}\bigg)^{2}\frac{\displaystyle \int_{
\lambda } \lambda F_{\lambda} S_{ \lambda } 10^{-0.4A_{\lambda}}
\textrm{d} \lambda }{\displaystyle \int_{ \lambda } \lambda f^{0}_{\lambda} S_{
\lambda } \textrm{d} \lambda }\bigg]+ZP_{S_{\lambda}}
\end{equation}
where $f^{0}_{\lambda}$ is a reference spectrum that gives a known apparent magnitude $ZP_{S_{\lambda}}$; in the \textsc{Vegamag} photometric system, which uses the flux of $\alpha$~Lyr as reference, $f^{0}_{\lambda}=F_{\lambda , {\rm VEGA}}$ and the zero-points $ZP_{S_{\lambda}}$ vanish according to our definition of WFI  \textsc{Vegamag} system (Paper I). We also assume an extinction $A_{\lambda}=0$, since we are interested in constructing a general observable grid of evolutionary models for unreddened objects. Eq.~(\ref{equation:first}) can be rewritten then as:
\begin{eqnarray} \label{equation:second} M_{S_{\lambda}} = &
-5\log\bigg( \frac{\displaystyle R_{\odot} }{\displaystyle 10 {\rm pc}
}\bigg) - 5\log\bigg(\frac{\displaystyle R_{\star}}{\displaystyle
R_{\odot}}\bigg) + B(F_{\lambda},S_{\lambda}) \nonumber \\ = & 43.2337 -
5\log\bigg(\frac{\displaystyle R_{\star}}{\displaystyle R_{\odot}}\bigg)
+ B(F_{\lambda},S_{\lambda})
\end{eqnarray}
where
\begin{equation}
\label{equation:third}
B(F_{\lambda},S_{\lambda})=-2.5\log\bigg(\frac{\displaystyle \int_{
\lambda } \lambda F_{\lambda} S_{ \lambda } \textrm{d} \lambda
}{\displaystyle \int_{ \lambda } \lambda F_{\lambda ,{\rm Vega}} S_{ \lambda
} \textrm{d} \lambda}\bigg)
\end{equation}
The latter term can be directly calculated having the synthetic spectrum for every \teff\ point of a given isochrone, a calibrated Vega spectrum and the band profile of the WFI photometric filter. The Vega normalization is performed using a recent reference spectrum of Vega \citep{bohlinvega}.

\begin{figure}
\epsscale{1.05}
\plotone{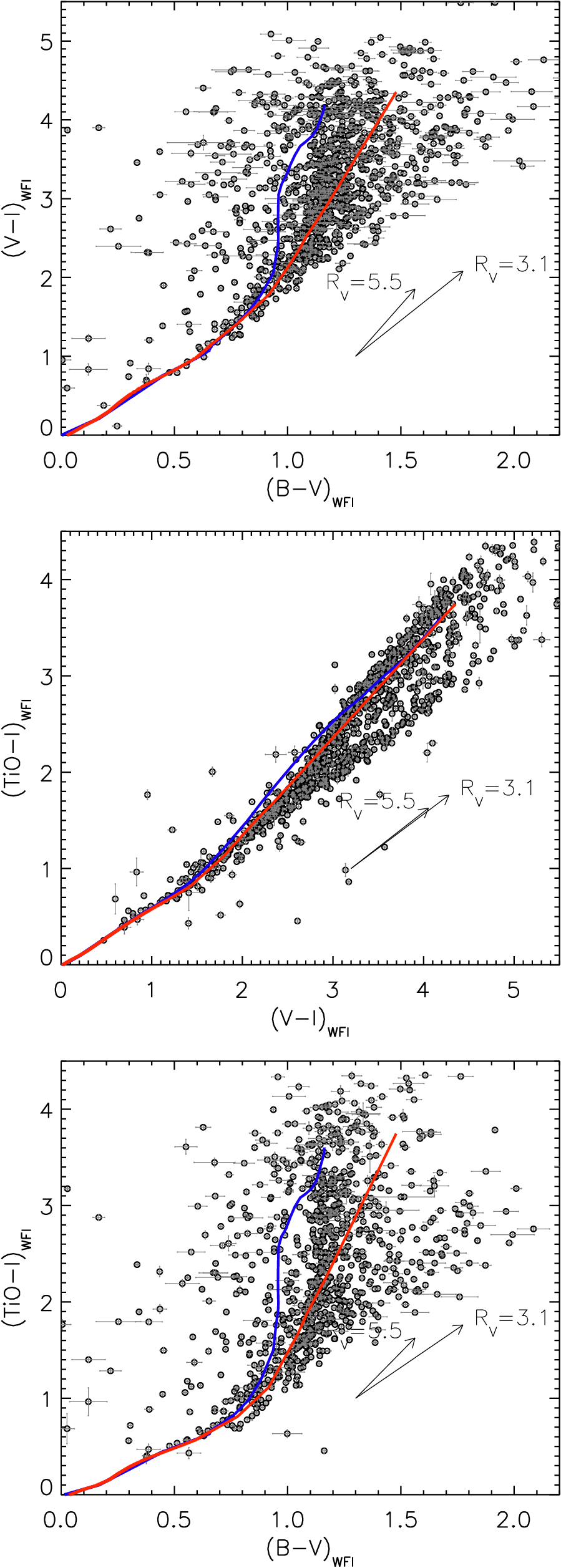}
\caption{Color-color diagrams for the ONC (circles). The solid lines are the loci of intrinsic photospheric emission computed by means of synthetic photometry, assuming a $2$ Myr Siess isochrone (blue line) and the Siess ZAMS (red line) to define $\log g$ as a function of \teff, for $A_V=0$. The reddening vectors corresponding to $A_V=2$~mag for two choices of $R_V$ are reported as a reference. \label{fig:colorcolor}}
\end{figure}

In Figure \ref{fig:colorcolor} we present three observed color-color plots for the ONC in $B$, $V$, $I$ and TiO bands, together with the intrinsic colors computed from synthetic photometry using our grid of spectra selecting $log g$ values from a \citet{siess2000} 2~Myr isochrone and for the zero age main-sequence (ZAMS) of the same family. The 2~Myr age is illustrative here, but consistent with the age of ONC derived from an HRD once \citet{siess2000} evolutionary models and a shorter distance for the ONC are used (though at least twice larger than the value found by H97 using \citet{dantona-mazzitelli94} evolutionary models). As shown in \citet{hillenbrand2008}, Siess models tend to predict higher ages, relative to other contemporary model isochrones, for a given PMS population in the HRD. Also, using a shorter distance for the ONC increases the age significantly from that derived by H97, by $\sim50\%$

In the same plot we show the extinction direction, for $A_V=2$. We have assumed two choices of the extinction law, the ``typical" Galactic one, described by a reddening parameter $R_V=A_V/(A_B-A_V)=3.1$ and the ``anomalous" one, considered more appropriate for the ONC \citep{johnson67,costero1970}, whose flatter variation with wavelength is described by a higher value $R_V=5.5$.
Since our photometry is expressed in the WFI instrumental photometric system, we computed the transformation of the reddening parameters $R_\lambda$ for the WFI bands as a function of $R_V$, which by definition refers to the Johnson photometric standard. We performed the computation, for several choices of the \emph{standard} reddening parameter $R_{V}=A_{V}/E(B-V)$ (in Johnson photometric system) applying the \citet{cardelli} reddening curve on our reference synthetic spectra, and measuring the extinctions in both Johnson and WFI photometric systems.
Examples of the transformation between $R_V$ (Johnson) and $R_V,R_I$\ (WFI) bands are listed in Table \ref{tablereddening}.

\begin{deluxetable}{ccc}
\tablecaption{Transformation between reddening parameters.}
\tablehead{\colhead{$R_{V}$} & \colhead{$R_I({\rm WFI})=\frac{A_I}{E(V-I)}$} & \colhead{$R_V({\rm WFI})=\frac{A_V}{E(V-I)}$} \\
\colhead{} & \colhead{} & \colhead{} }
 \startdata
2.0 & 0.69 & 1.69 \\
3.1 & 1.06 & 2.06 \\
4.3 & 1.31 & 2.31 \\
5.5 & 1.49 & 2.49 \\
6.0 & 1.53 & 2.53
\enddata
\label{tablereddening}
\end{deluxetable}

A broad range of optical colors is exhibited by the ONC stars in Figure \ref{fig:colorcolor}. Stars to the far right of the model ZAMS lines are heavily reddened stars, while stars to the far left, especially along $B-V$ axes, can be explained in terms of highly accreting stars, as we demonstrate in Section \ref{accretioncmd}.  Stars of intermediate color require detailed interpretation.  It is important to recall from above that ZAMS colors are not necessarily applicable to the young stellar population of the ONC.  According to the models, the photospheric colors of pre-main sequence stars should vary significantly with age, in particular in $BVI$ colors.  From Figure \ref{fig:colorcolor} it is evident that the model ZAMS is systematically redder in $B-V$ than much our data, in a way that is unexpected from a distribution of accretion rates.  The effect is significant for large colors ($V-I\gtrsim2$), corresponding to \teff$\lesssim4000$~K, or M-type stars.  Apparently the ONC color sequence is the locus seen just to the left of the model ZAMS.   This suggests either a problem with the ZAMS relation or that the intrinsic colors of young pre-main sequence stars are bluer than main-sequence dwarfs.  We now explore whether the atmosphere models we have selected are appropriate to describe the intrinsic emission of first the ZAMS and then the young ONC stars.

We validate that the models can reproduce the observed colors of main sequence stars by showing
in Figure \ref{fig:ZAMS_comparison_empirical_synthetic} the 2Myr
model and the ZAMS model computed in the Johnson-Cousins $BVI_c$ bands
using our grid of synthetic spectra, as well as empirical colors valid of
MS dwarfs, from \citet{kenyon-hartmann95} and \citet{vandenberg2003}.
We test the correctness of the model ZAMS in the standard Johnson-Cousins bands
since we don't have at disposal an empirical color scale for MS dwarfs in the WFI
phtometric system. We stress that due to the differences between Johnson-Cousins and WFI systems,
the predicted colors are slightly different than those presented in Figure
\ref{fig:colorcolor}a, in particular the $(B-V)$ color is larger than that
for our WFI bands; however the qualitative behavior remains unchanged:
a young isochrone is significantly bluer than the ZAMS. On the other hand
the model ZAMS is compatible with empirical color scales derived for the same
photometric system, with offsets much smaller than the difference between
the 2~Myr locus and the modeled ZAMS.  This confirms that the ZAMS computed
by means if synthetic photometry on our grid of atmosphere model
is fairly accurate.

\subsection{Constraining the intrinsic colors for ONC stars}
\label{section:emprical_colors}

Returning now to the WFI photometric system and Figure \ref{fig:colorcolor},
the question becomes what is the appropriate set of intrinsic photospheric colors to utilize.
In general, the non-uniform extinction of the members displaces each star in the direction
of a reddening vector, spreading the location of any hypothetical ONC isochrone in color.
We assume that the reddening distribution is similar for members having different spectral types; this is a fairly good approximation but for the lowest intrinsic luminosities for which high $A_V$ are below the detection threshold. Thus, the fiducial points of the observed distributions for the ONC population are expected to be parallel to any appropriate pre-main sequence model computed for $A_V=0$, and located at redder colors than the intrinsic photospheric locus (i.e. shifted by a non-negative amount of extinction).

Although reddening from a 1 or 2 Myr isochrone could explain the observed displacement of the
observed color locus redward of the model isochrones but blueward of the ZAMS,
it is not adequate to simply adopt a model isochrone since one of the goals
of this study is to assess the age and age spread for the ONC population.
It is instructive to study the constraints available from our own photometry
on the actual intrinsic $BVI$ colors that are appropriate.
We define an empirical intrinsic color isochrone in $BVI$ for stars in the ONC by considering a set of stars
that are free of both blue excess due to accretion and reddening due to extinction.

\begin{figure}
\epsscale{1.1}
\plotone{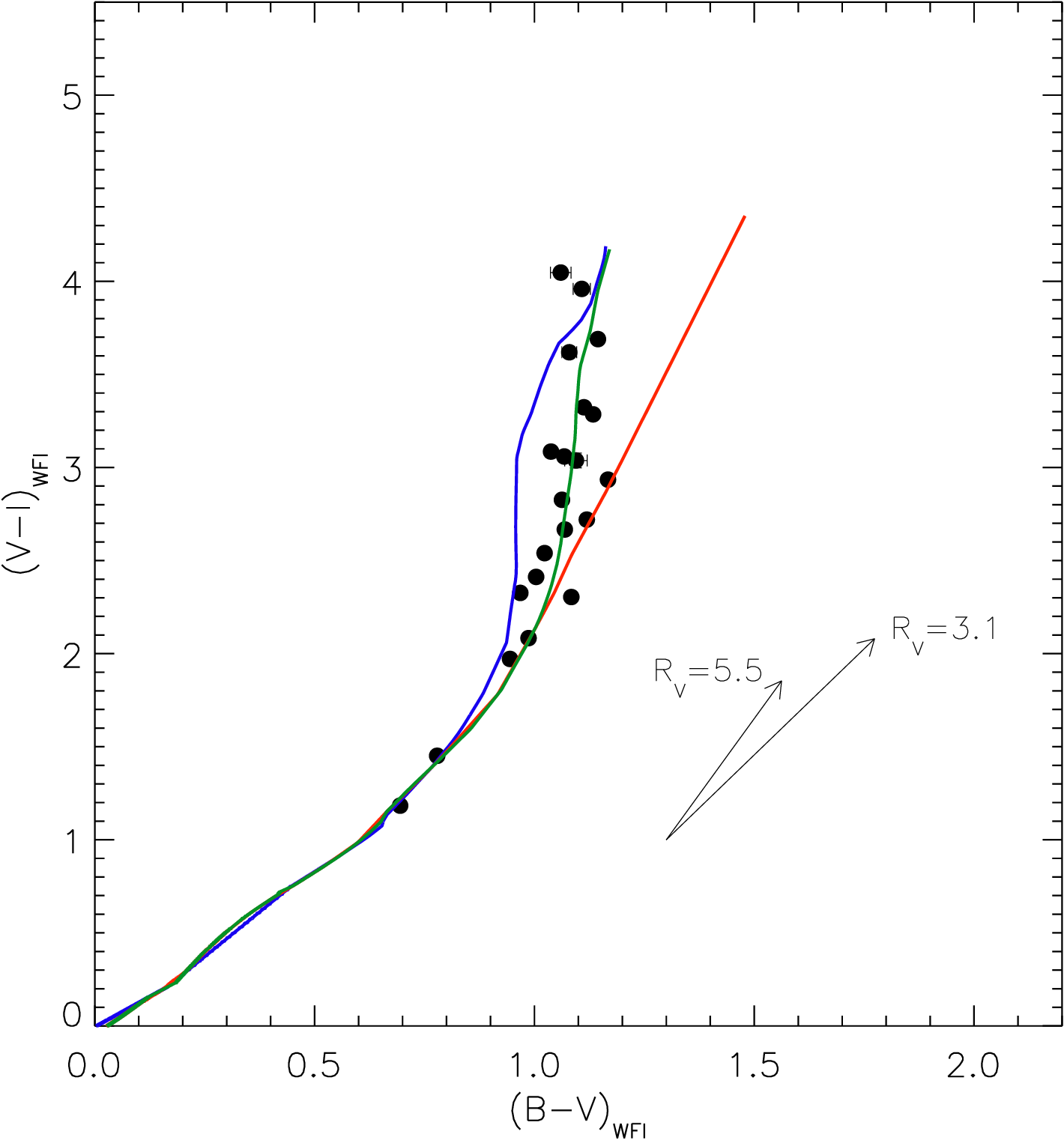}
\caption{$BVI$ color-color plot including only sources with small photometric error ($\leq 0.1$~mag), membership probability known and $\geq98\%$, no H$\alpha$, excess, modest extinction from H97 ($0.1<A_V<0.5$) (dots). The data have been de-reddened based on the extinctions measured from H97. As in Figure \ref{fig:colorcolor}, the red and blue lines represent respectively the synthetic ZAMS and 2~Myr model. The green line is our empirical calibration of the intrinsic colors, intermediate between the two (see text).
\label{fig:colorcolorempirical}}
\end{figure}
From Figure \ref{fig:colorcolor}a,c it is evident that the 2~Myr synthetic colors, when reddened, are more compatible with the observed sequence than the ZAMS locus. However, it is hard to provide a quantitative estimate of this level of agreement. This is because the observed sequence shows a large scattering, mainly due to the differential extinction but, as we will demonstrate in Section \ref{accretioncmd}, also because of broad-band excesses caused by stellar accretion.
To simplify comparison of the model colors with our data, we isolate stars that:
\begin{itemize}
  \item are candidate non-accreting objects: we utilize the H$\alpha$ excesses presented in Paper~I and we select the sources with no excess.
  \item have modest extinction: we consider the extinctions from H97, and select sources with $0<A_V<0.5$
  \item are known ONC members: we consider the memberships tabulated in H97, mainly collected from \citet{jones-walker88}, and select stars with membership probability $m \geq 98\%$
  \item have small photometric errors: we restrict errors to $<0.1$~mag in all $BVI$ bands.
\end{itemize}
Restricted in this way, the sample satisfying all of the above criteria contains $\sim21$ stars and includes the known ONC members for which the observed fluxes are closest to simple photospheres. By dereddening their colors subtracting the modest extinctions measured in H97, these stars should lie on our color-color diagram along a sequence which represents the intrinsic colors valid for the ONC. This color locus is illustrated in Figure \ref{fig:colorcolorempirical}, where we also overlay the predicted intrisic colors from synthetic photometry for the ZAMS and the 2~Myr isochrone.
While the intrinsic colors of this sample of non-accreting ONC stars are bluer than those along the ZAMS,
they are not as blue as the theoretically predicted colors for a 1-2 Myr population.

We thus confirm that the intrinsic colors for the ONC are not in agreement with the dwarf colors, but rather bluer as qualitatively predicted by the 2~Myr isochrone. However, small offsets are still present with respect to the young isochrone that are color-dependent.  In particular, while the 2~Myr model seems appropriate for the coolest stars ($\sim3000$~K, corresponding to $(V-I)\sim4$), the observed separation of the ONC color sequence from the ZAMS locus occurs at slightly larger colors ($V-I\gtrsim2$) than predicted by the 2~Myr model. We chose therefore to use the empirical data to refine the model isochrone colors so as to best match the data of Figure \ref{fig:colorcolorempirical}. To this purpose we consider the ZAMS intrinsic colors for \teff$\geq3700$~K, the 2~Myr synthetic colors for \teff$\leq2900$~K, and a linear interpolation between the two for $2900$~K$<$\teff$<3700$~K. The result is shown also in Figure
\ref{fig:colorcolorempirical}, green line; it is evident how this empirical correction of the intrinsic colors now allows for a good match to the selected sample of non-accreting non-extinguished members of the ONC.

We refer hereafter as the {\em reference model} these empirically-validated intrinsic $BVI$ colors, which can be considered a fairly good approximation of the photospheric fluxes, as a function of \teff, for the ONC.

\subsection{Bolometric corrections}
\label{section:bolometric_correction}
As described above, intrinsic colors for 2~Myr and ZAMS stars were derived by means of synthetic photometry performed on a grid of spectra using the appropriate $\log g$ vs. \teff.  For the intermediate $\log g$ values between the two models that is needed for our empirical isochrone, we adopt ZAMS values for \teff$\geq3700$~K, 2~Myr for  \teff$\leq2900$~K, and a linear interpolation in between) to provide the empirically constrained set of model
spectra that match the optical colors of the ONC.

We can use these {\em reference spectra} to also compute the bolometric corrections (BC) valid for our filters without relying on relations taken from the literature that, besides requiring color transformations, may not fully apply to pre-main-sequence stars. Clearly, based on our data, we cannot validate the accuracy of the BCs we derive from synthetic photometry. However, based on the evidences we have presented in Section and \ref{sectionsyntheticcmd} and \ref{section:emprical_colors} suggesting that our reference model is adequate to describe the photospheric emission of the ONC stars at optical wavelengths, we are confident that the BCs derived on the same spectra are similarly appropriate.

The bolometric corrections are defined as the difference between the bolometric magnitude of an object, found by integrating the SED across all wavelengths, and the magnitude in a given band:
\begin{equation}
\label{BCequation}
{\rm BC} = 2.5\cdot \log\bigg[\frac{\int f_\lambda (T)\cdot S_\lambda d \lambda}{\int f_\lambda (T) \cdot d \lambda}\bigg/ \frac{\int f_\lambda (\odot)\cdot S_\lambda d \lambda}{\int f_\lambda (\odot)\cdot d \lambda}\bigg] +C,
\end{equation}
where $C$ is a constant equal to the bolometric correction for the Sun, BC$_\odot$,  in that particular photometric band.  The values of $C$ are irrelevant in our case, since for obtaining the total luminosities in terms of $L/L_\odot$ (see Section~\ref{sectionH-R} and Eq. \ref{eq:logl_loglsun}), BC$_\odot$ is subtracted and vanishes; thus we impose $C=0$. The integrals in Eq.~\ref{BCequation} have been computed for the $V$ and $I$ bands for the  reference spectra, and for consistency we considered as a reference solar spectrum a NextGen SED interpolated on the grid for solar values with \teff=5780 K, $\log g = 4.43$, and $\MH=0.0$.

\begin{figure}
\epsscale{1.1}
\plotone{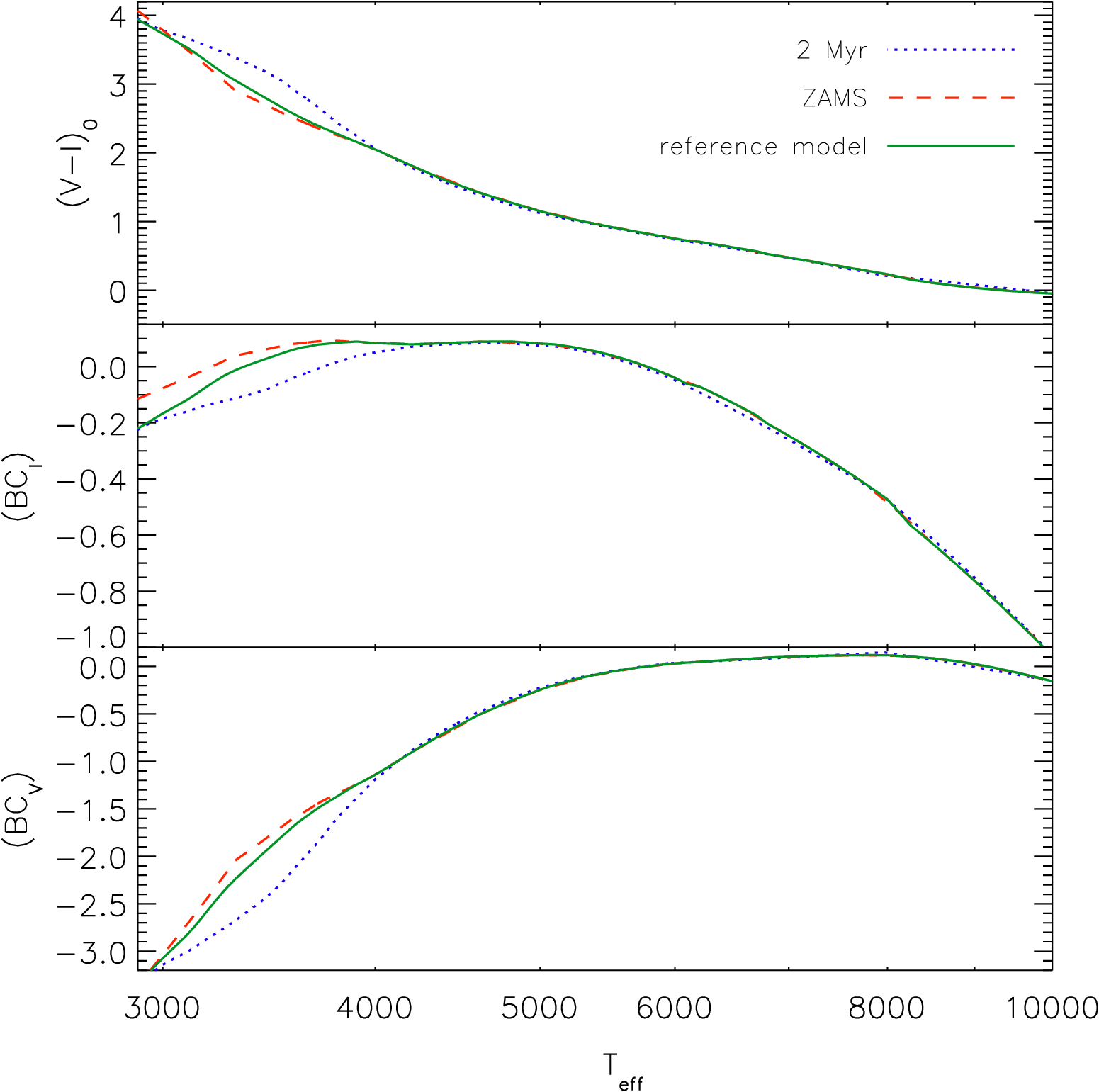}
\caption{\emph{Upper panel:} intrinsic color $(V-I)_0$ as a function of temperature, computed for our 2~Myr isochrone, the ZAMS, and our empirically calibrated reference model, in the WFI instrumental bands.
 \emph{Middle panel and lower panels:} Bolometric corrections derived in the same way, for the $I$ and $V$ WFI instrumental bands.
    \label{figbc}}
\end{figure}

The results are shown in Figure \ref{figbc} as a function of temperature. For comparison, we also show these relations computed in the same way but using the surface-gravity values of the Siess ZAMS and the Siess 2~Myr isochrone. As for the colors, important differences arise for \teff$<4000~K$, corresponding to M-type stars. In these regime, the difference between the three models is up to 0.15 mag for BC$_I$ and up to 0.6 mag for BC$_V$. For the derivation of the H-R diagram (see Section \ref{sectionH-R}) we utilize only BC$_I$, and the measured difference of 0.15 mag between the 2~Myr model and the ZAMS translates into an offset of 0.06~dex in $\log L$, amount small enough to be irrelevant for the computation of the stellar parameters. Therefore, we assume the reference model BCs for consistency with the model we use to derive the intrinsic colors, confident that possible uncertainties in the BCs do not affect significantly the result. The major advantage of our approach is that we can compute consistently these quantities for the WFI photometric system which our photometry follows.

\subsection{Extinction law}
\label{section:reddening}

\begin{figure}
\epsscale{1.1}
\plotone{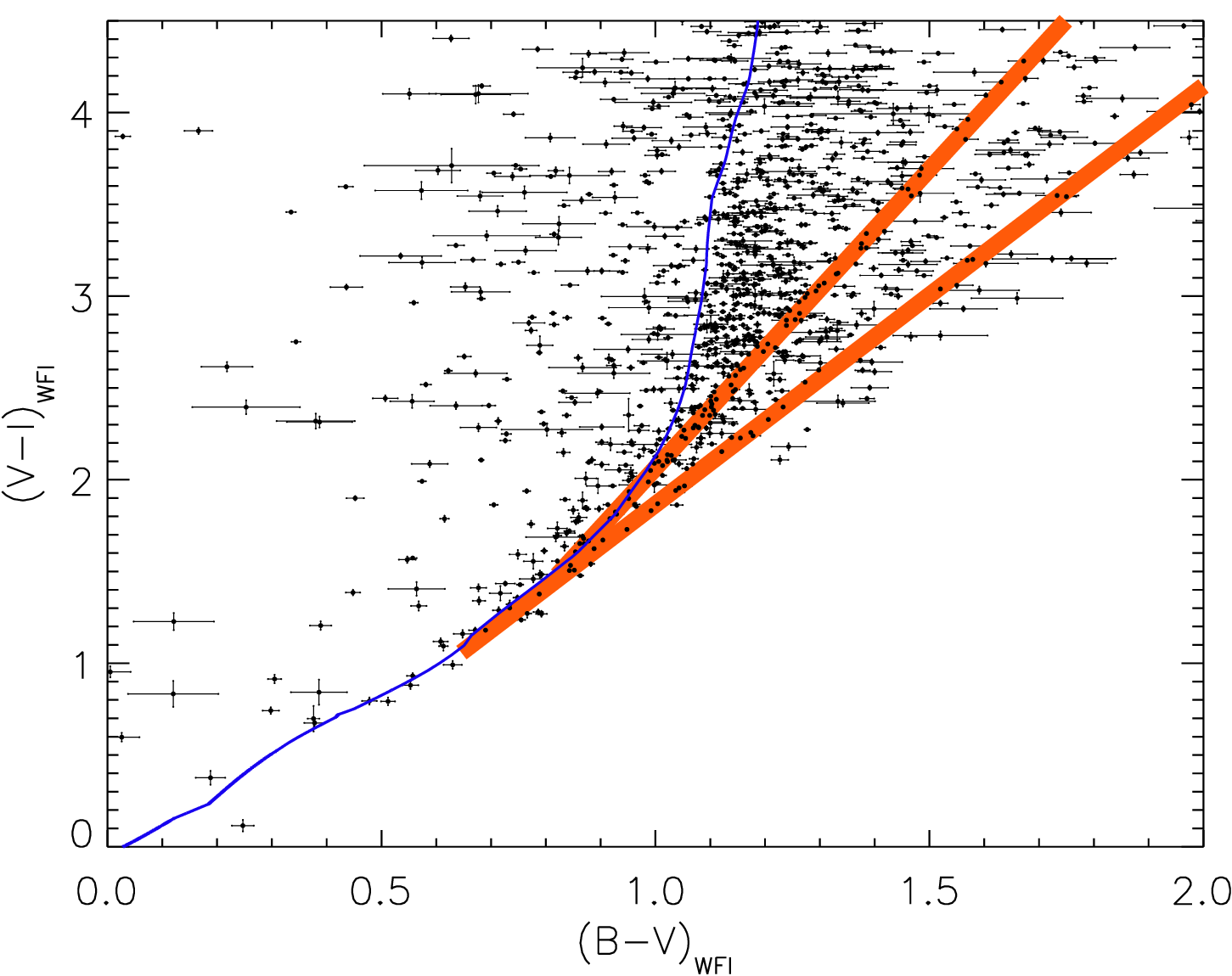}
\caption{($B-V$) vs ($V-I$) plot of the ONC population (dots), with the 2~Myr reference model (solid line). The two thick lines correspond to the reddening direction assuming $R_V=5.5$ (upper line) and $R_V=3.1$ (lower line) extinction laws, applied to the tangent points of the model.  \label{fig:colorcolor_rv}}
\end{figure}

We are making an implicit assumption that the extinction in the entire region under study follows the same reddening law. Differences in  grain size and composition might be expected, due to the complex structure of the ISM within the Orion Nebula, the non-uniformity of the foreground lid of neutral material and, for a fraction of members, the presence of circumstellar disks in different evolutionary state. Optical photometry alone cannot provide an independent assessment of the most adequate reddening law for each source, given the degeneracy for late-type stars between the observed colors and \teff\ and the presence of non-photospheric effects. From our color-color plots, however, we can make some general considerations. Whereas the $(V-I)$ vs. $(TiO-I)$ plot of Figure \ref{fig:colorcolor} does not provide useful information, as  the reddening vectors for the two choices of $R_V$ are nearly parallel due to the limited wavelength range spanned,  in the $(V-I)$ vs. $(B-I)$ plot the two reddening vectors are clearly separated. For spectral types $K$ and earlier ($(V-I)<2$), our template model is nearly parallel to the two reddening vectors (i.e., the above mentioned degeneracy between spectral types and reddening), meaning that all the highly extinguished members within this range should be located in a narrow stripe at higher color terms.

In Figure \ref{fig:colorcolor_rv} we show the reddening directions for the two values of $R_V$ applied to the tangent points with the reference model. If $R_V=5.5$ were the correct choice, no stars would be detected to the right from the associated reddening direction, in contrast with our data. On the contrary, the $R_V=3.1$ reddening limit encompasses almost the entire photometry, with the exception of few scattered stars with high photometric errors in $(B-V)$. Clearly, the scatter in our data and the photometric errors limit a precise estimate of the actual value of $R_V$. Nevertheless, we can conclude in an unquestionable way that the typical galactic law is more compatible with our data than the \citet{costero1970}.

Given the uncertainties, and for the purpose of discussing how the choice of $R_V$ affects the scientific results, we perform the analysis presented in the next sections using both values, but we present mostly the results for $R_V=3.1$, mentioning any systematic changes induced by assuming the higher $R_V$, where relevant.

\subsection{Mass accretion}
\label{section:accretionCMD}
The color-color diagrams of Figure \ref{fig:colorcolor} show that there is a fraction of stars lying significantly to the left of the models. In this region, according to our reference model, a star of given temperature and positive reddening cannot end up. The fact that the TiO-I vs V-I frame is the ``tightest'', while the ($TiO-I$) and ($V-I$)  vs. ($B-V$) are more scattered suggests that these sources have significant excess in the B band. Moreover, low ($V-I$) stars are scattered only bluewards with respect to the reference model, confirming the presence of a $B$-band excess for a subsample of sources with low reddening.

We investigated the possibility that the blue excess is due to accretion luminosity. According to the current paradigm for classical T Tauri stars (CTTS), gas is accreted onto the central star from an accretion disk  truncated at a few stellar radii by the stellar magnetic field. From the inner edge of the disk, the material falls along infall columns beamed by magnetic field lines \citep{koenigl91, shu94, muzerolle98b}.  Several lines of observational evidence support the magnetospheric model, including: (1) the broad emission lines emitted by the infalling gas, presenting occasionally redshifted absorption components \citep{calvethartmann92, muzerolle98a}; (2) the near-infrared spectral energy distributions (SEDs), consistent with those of disks truncated at small radii \citep{kenyon94, meyer97}; (3) the variations in luminosity consistent with the presence of \emph{hot spots}, originating where the accretion flow impacts the stellar surface \citep{herbst94, gullbring2000}.
The presence of ongoing accretion processes in the ONC has been investigated by several studies, e.g., in \citet{rebull2000} and \citet{robberto04} from the $U$-band excess is used to obtain mass accretion rates, or in \citet{sicilia-aguilar2005} and \citet{hillenbrand98disks} using optical spectroscopy.

\begin{figure}
\epsscale{1.1}
\plotone{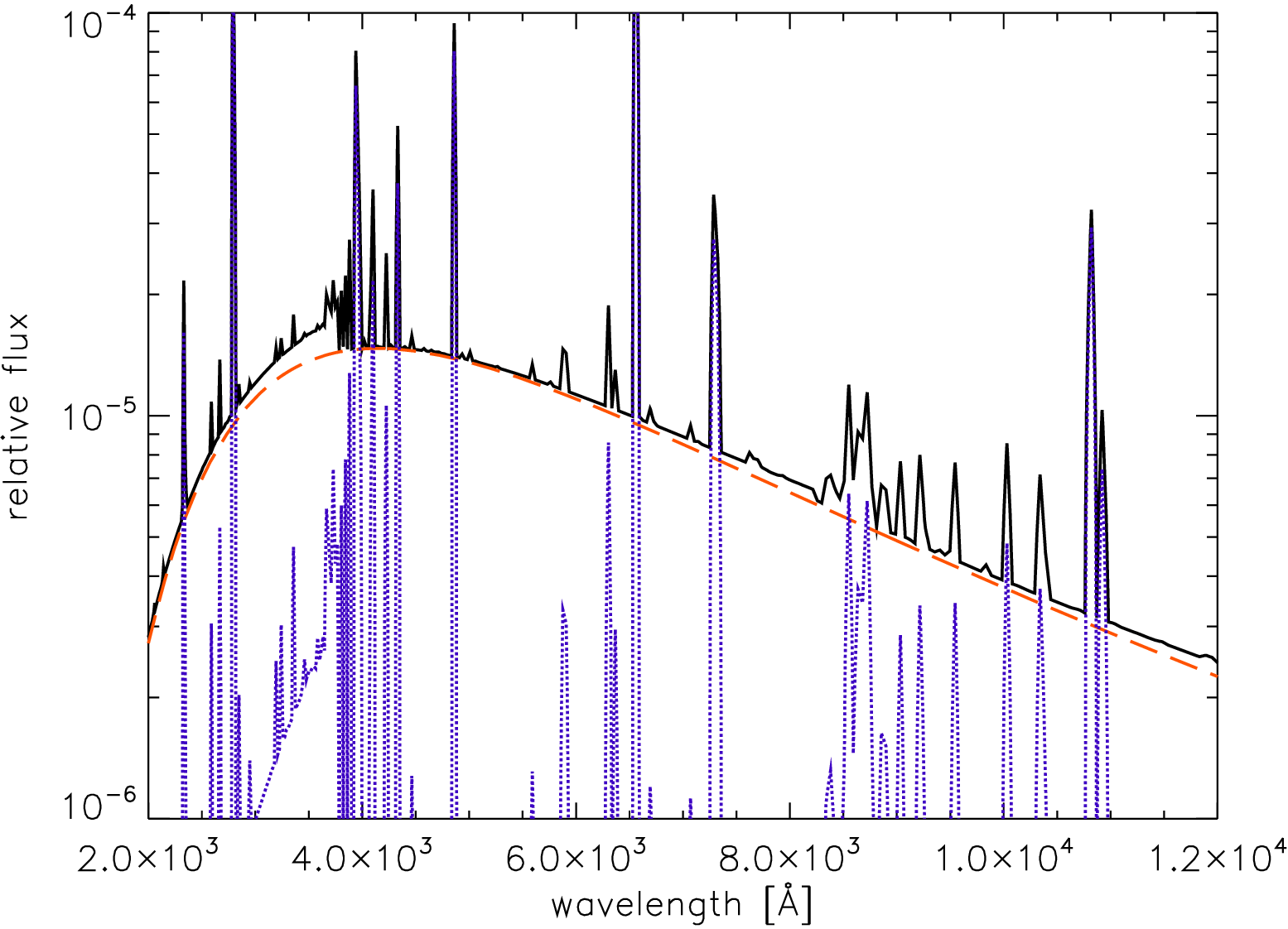}
\caption{Accretion spectrum simulated with \textsc{Cloudy}. The solid line is the total emission, which is the superposition of an optically thick emission, with \teff$=$7000~K, of the heated photosphere (\emph{dashed line}) and  the optically thin emission of ionized gas with density $n=10^8$~cm$^{-3}$ (dotted line). \label{accretionspectrum}}
\end{figure}

In order to verify the possible role of accretion also in our $BVI$ bands, we simulated a typical accretion spectrum to be added to our reference models. According to \citet{calvetgullbring98}, the total accretion column emission can be well approximated by the superposition of an optically thick emission from the heated photosphere below the shock, and an optically thin emission generated in the infalling flow. The SED of the optically thick emission is compatible with a photosphere with \teff\ = 6000-8000~K, while the optically thin pre-shock column can be modeled by thermal recombination emission. The relative fraction of the two components with respect to the total emission produced by accretion is about $3/4$ and $1/4, $ respectively. We therefore assumed for the optically thick emission a black body with $T=7000$~K, whereas for the thin emission we used the version $07.02.01$ of the {\sc Cloudy} software, last described by \citet{ferland98}, to reproduce the spectrum of an optically thin slab ($n=10^8$~cm$^{-3}$) also at $T\simeq7,000$~K.
The derived, low resolution, accretion spectrum is shown in Figure \ref{accretionspectrum}.

\begin{figure}
\epsscale{1.1}
\plotone{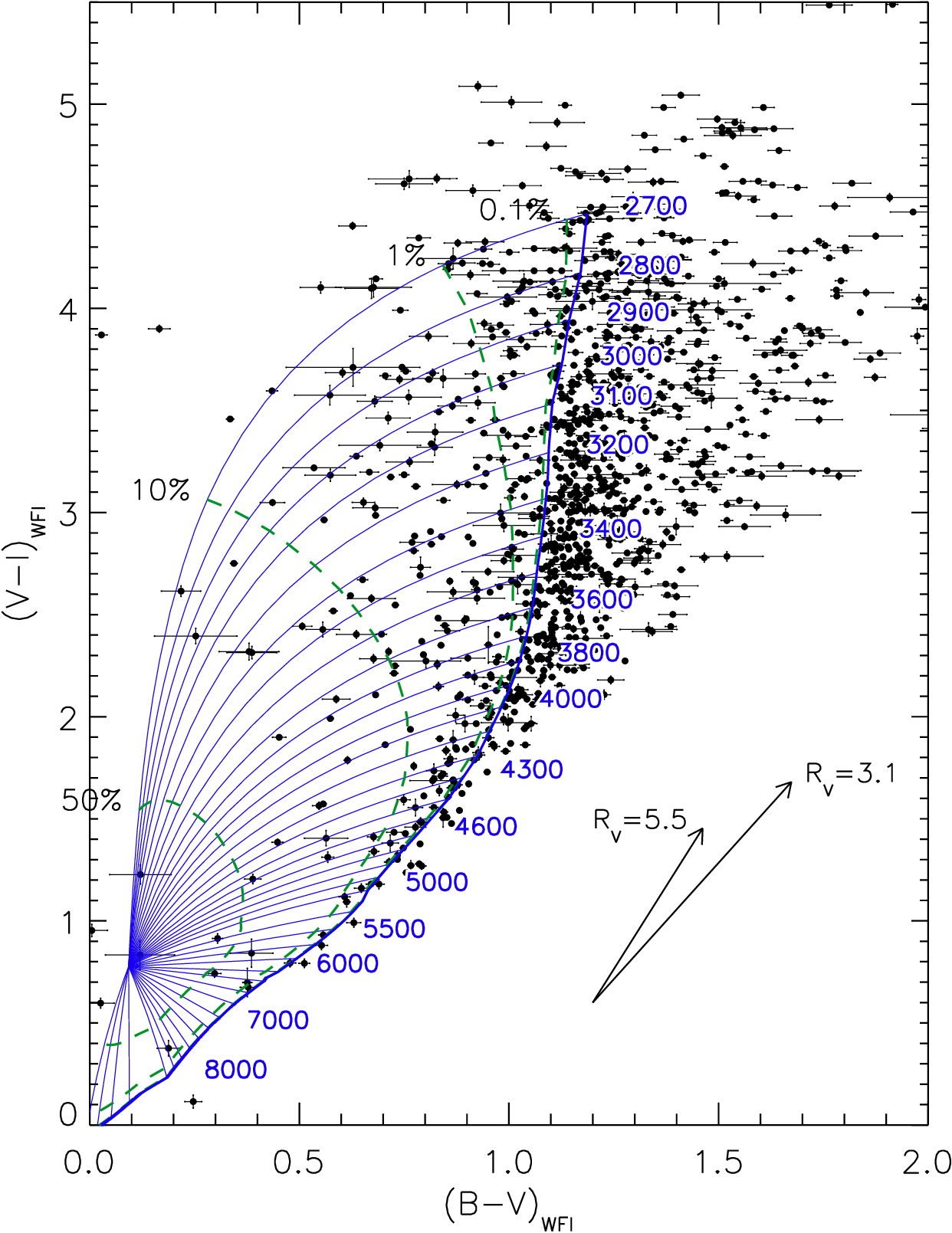}
  \caption{$BVI$ color-color diagram of our WFI catalog. The thick line represents the synthetic CMD obtained from the reference spectra selected (see sub\S\ref{section:emprical_colors}) for $A_{V}=0$ and the pertinent effective temperatures are shown along the curve. Thin blue lines represent the computed displacement, for a star of a given temperature, adding a component of the accretion spectrum (as in Figure \ref{accretionspectrum}) and increasing the parameter $L_{\rm accr}/L_{\rm tot}$ from $0$ to $1$. Loci for $L_{\rm accr}/L_{\rm tot}=0.5$, $0.1$, $0.01$ and $0.001$ are overplotted (\emph{dashed lines}). Reddening vectors corresponding to an extinction of $A_V=2$ are shown for the two choices of reddening parameter $R_V$ assumed in this work. \label{accretioncmd}}
\end{figure}
We calculated the displacement in the color-color diagrams caused by the presence of ongoing accretion by adding these spectra, with an increasing ratio of $L_{\rm accr}/L_{\rm tot}$, to our reference model. For any given combination, we recalculated the synthetic photometry, deriving  the  $(B-V)$, and $(V-I)$ colors in the WFI instrumental photometric system.
The results (Figure~\ref{accretioncmd}) show the derived loci for each stellar  temperature, all converging to the point representative of the pure accretion spectrum, as well as for fixed  $L_{\rm accr}/L_{\rm tot}$ ratios.

The blue excess that is observed for a number of the ONC targets in the $(V-I)$~vs.~$(B-V)$ diagram apparently can be explained by an accretion spectrum added to the stellar emission, with typical luminosity well below $10\%$ of the total luminosity of the star.
Figure \ref{accretioncmd} clearly shows that small fractions of accretion luminosity have a larger effect on the colors of low-temperature stars, indicating that accretion effects must be taken into account especially in the study of the low-mass end of the ONC.

\begin{figure}
\epsscale{1.1}
\plotone{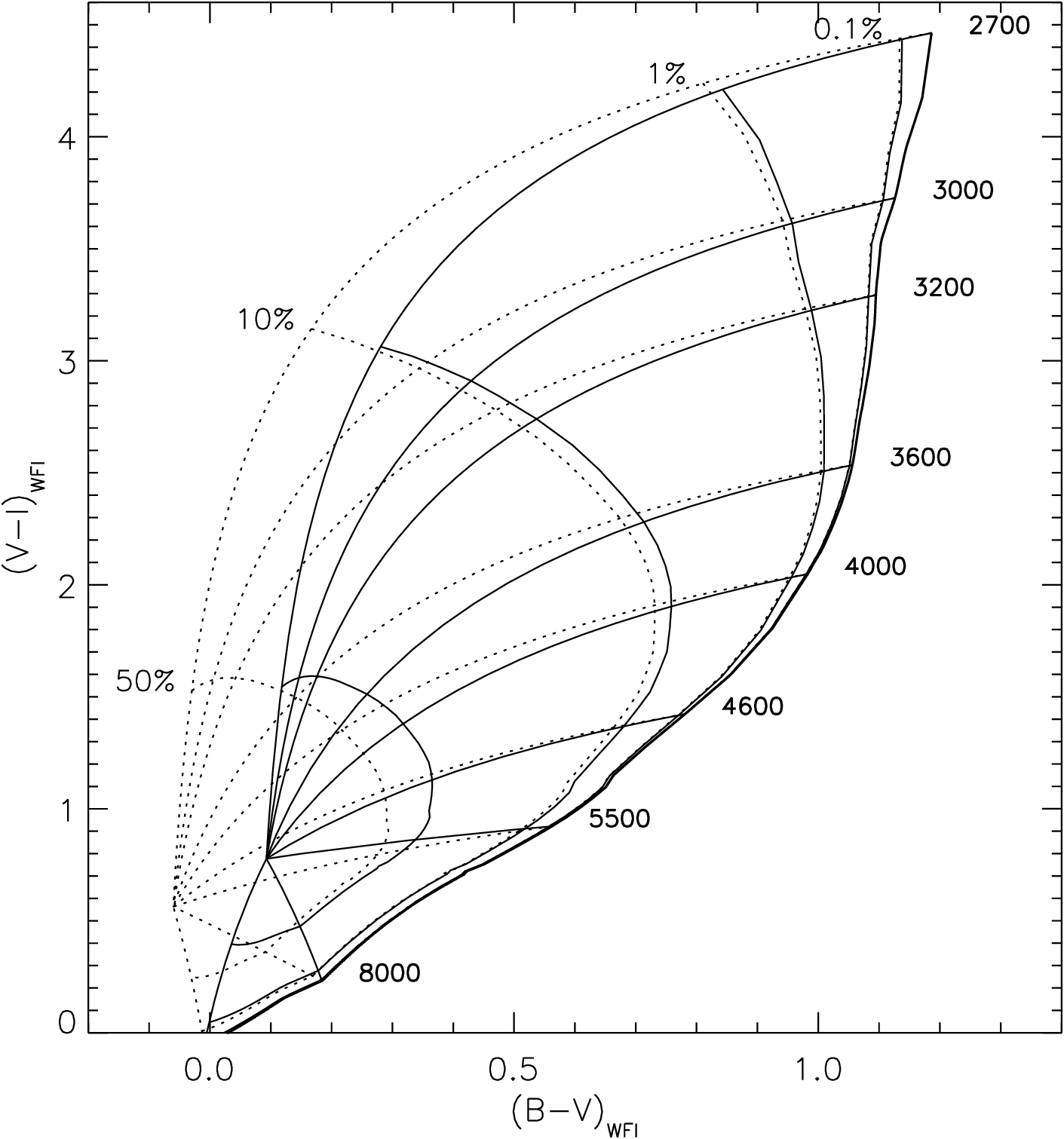}
  \caption{Same as Figure \ref{accretioncmd}, showing the difference in modeled displacement caused by accretion increasing the accretion temperature from 7000~K (assumed in the case of Figure \ref{accretioncmd}, solid lines) to 9000~K (dotted lines). \label{accretioncmd_comparison}}
\end{figure}

We have tested how critical is the assumption of a unique accretion spectrum for the modeling of the color excess. When the accretion temperature is set to be 9000~K instead of 7000K (see Figure \ref{accretioncmd_comparison}), the colors of the pure accretion spectrum (i.e. the location, in the diagram of Figure \ref{accretioncmd} where all the curves join) decrease by 0.15 mag in $(B-V)$ and 0.2 mag in $(V-I)$. This is a small difference already in the limiting case of $L_{\rm accr}/L_{\rm tot}$=1, for much lower accretion, which is the case of the majority of our stars, the position of the accretion curves of Figure \ref{accretioncmd} remains almost unchanged. We conclude  that our approximation of applying a unique accretion spectrum to model the stellar colors contamination is enough robust for our purposes.

\section{Analysis of stars with known spectral type}
\label{section:subsampleanalysis}
In the previous sections we have isolated the intrinsic color scale which best fits the photospheric emission of the ONC members. From  the intrinsic colors, taking also into account possible color anomalies through the addition of an accretion spectrum, we can determine the extinction  $A_V$ and therefore the stellar luminosity.
The combination of spectroscopic information with accurate photometry, together with our determination of the intrinsic relations (intrinsic colors and BCs for the young ONC, and the treatment of $L_{\rm accr}$ excess), allow us to build the H-R diagram for the ONC improving over the analysis of previous works.

\subsection{Spectral types}
\label{subsection_spectraltypes}
We take full advantage of the catalog of spectral types in the ONC of H97 to determine the temperatures of as many members as possible.  Matching our catalog with the H97 spectroscopy (see Paper I), we isolated 876 sources with available spectral type.

We extend the H97 sample of spectral types using new spectroscopic data.
The observations were obtained as part of a follow-up to the study of \citet{stassun99} of the larger Orion Nebula star forming region.
The spectra were obtained with the WIYN 3.5m multi-object spectrograph at Kitt Peak National Observatory on UT 2000 Feb 11. The instrument setup was the same as described in H97, providing wavelength coverage over the range $5000$\AA$ < \lambda < 9000$\AA\  with spectral resolution of $R\sim1000$. Exposure times were 2 hours and the typical signal to noise ratio was $\sim50$ per resolution element. Spectral types were determined from the spectra using the same spectral indices described by H97. In particular, for our targets which are principally of K and M type, we found the TiO indices at $6760$\AA, $7100$\AA, and $7800$\AA\  to be most useful. We calibrated the indices against observations of stars previously classified by H97 as well as with the template library of \citet{allen-strom95}. For 85\% of the sources a spectral type is assigned with a precision of 1 subtype or less.
From these observations, we include the 65 stars that overlap with the WFI FOV.

We also add 182 stars classified in Paper~I from the TiO spectro-photometric index (hereafter [TiO]). These sources are limited in the spectral range M1-M6. Earlier spectral types are not classified with this method because our index correlates with a TiO absorption band present only in M-type stars. On the other hand, for late M-type stars we were limited by the detection limit of the 6200\AA\ narrow-band filter. Given the small range of \teff\ spanned, this selection effect in \teff\ corresponds to a selection in masses, since the evolutionary tracks are nearly vertical in the HRD.

Thus, spectral classification is available for 1123 out of 2621 stars with optical photometry. For 121 of them, only $I-$band ground-based photometry is available, and we exclude them from our analysis since the lack of at least one color term makes the derivation of the reddening impossible. The sample of stars with known spectral type and at least $V$ and $I$ magnitudes accounts therefore for 1002 objects, 820 of them with slit spectroscopy (from both H97 and our new spectroscopic survey) and 182 M-type stars classified by means of [TiO] index.

For the stars also present in the H97 catalog, we utilize the membership probability collected in that work, which is taken from the proper motion survey of \citet{jones-walker88} and assigning a $99\%$ probability to the externally photo-evaporated disks (\emph{proplyds}, \citealt{bally2000,ricci2008}) located in the vicinity of the Trapezium cluster. Among our 1002 stars, 703 are confirmed members, with a membership probability $P>50\%$, while 54 are non-members ($P<50\%$). The distribution of memberships is bimodal, peaking at $P=99\%$ and $P=0\%$. This implies that our choice to define the limit at 50\% does not influence significantly the selection. In fact selecting only sources with $P>80\%$ results in selecting only 12 fewer stars.
For 245 the membership is unknown, but given the relatively low fraction of confirmed non-members, we expect all but 20-30 of this stars to be ONC members. This contamination, corresponding to $2$--$3\%$ of the sample is too small to bias our results; therefore we consider the unknown membership stars as \emph{bona fide} members, taking advantage of the inclusion of $\sim35\%$ more stars in our sample to improve the statistical significance of our results.

\subsection{Analysis of completeness}
\label{section:completeness_LAH}
In paper I we obtained the completeness function of our photometric catalog, based on the comparison of the luminosity functions (LFs) derived for WFI with that of the  HST/ACS survey on the ONC, characterized by a much fainter detection limit. We refer to this as the {\em photometric completeness}.
Here we compute the completeness function of the sample of stars that have at least both $V$ and $I$ magnitude in our catalog and and available spectral type. We refer to this one as the HRD {\em subsample completeness.}

The spectroscopic catalog of H97 is shallower than our photometry, with a 50\% completeness at $I_c\simeq 17.5$ mag, and our extension from new spectroscopy does not change significantly this limit\footnote{The new spectra we have presented do not extend the H97 spectral catalog to fainter magnitudes, and are mostly located outside the FOV of H97, in the peripheral regions of our WFI FOV.}.
On the other hand, the subsample of stars classified with the [TiO] index is about 0.5~mag deeper than in H97, to $I\sim18$~mag and does not cover the bright end of the cluster, populated by earlier spectral types (see Paper~I).

The $V-I$ color limit is essentially set by the photometric completeness at $V-$band. Our $I-$band photometry is much deeper than the $V-$band: all the sources detected in $V$ are also detected in $I$ whereas we detect twice as much sources in $I$ than in $V$. On the other hand $B-$band is shallower than $V$, but to provide an estimate of the stellar parameters at least one color (i.e., $(V-I)$) is needed (see Section \ref{section:ReddAcc}). As a consequence, the inclusion of a star in the subsample used to derive the stellar parameters, is limited by only a) \teff\ estimates; b) $V-$band photometric detection.

The $V-$band \emph{subsample completeness} function is the product of the \emph{photometric completeness} in $V$ and the fraction of sources, as a function of $V$ magnitude, that have a \teff\ estimate.
We computed the latter simply as a ratio between the LFs in $V-$band of the subsample and the entire catalog.
In this way the subsample completeness accounts for both the detection limit of the photometry, and the selection effects from the non-homogeneity of the sub-sample with \teff\ estimates.  In other words, it represents the probability of a star of a given $V-$magnitude to be detected in $V$ and $I$ and to be assigned a spectral type.

\begin{figure}
\epsscale{1.1}
\plotone{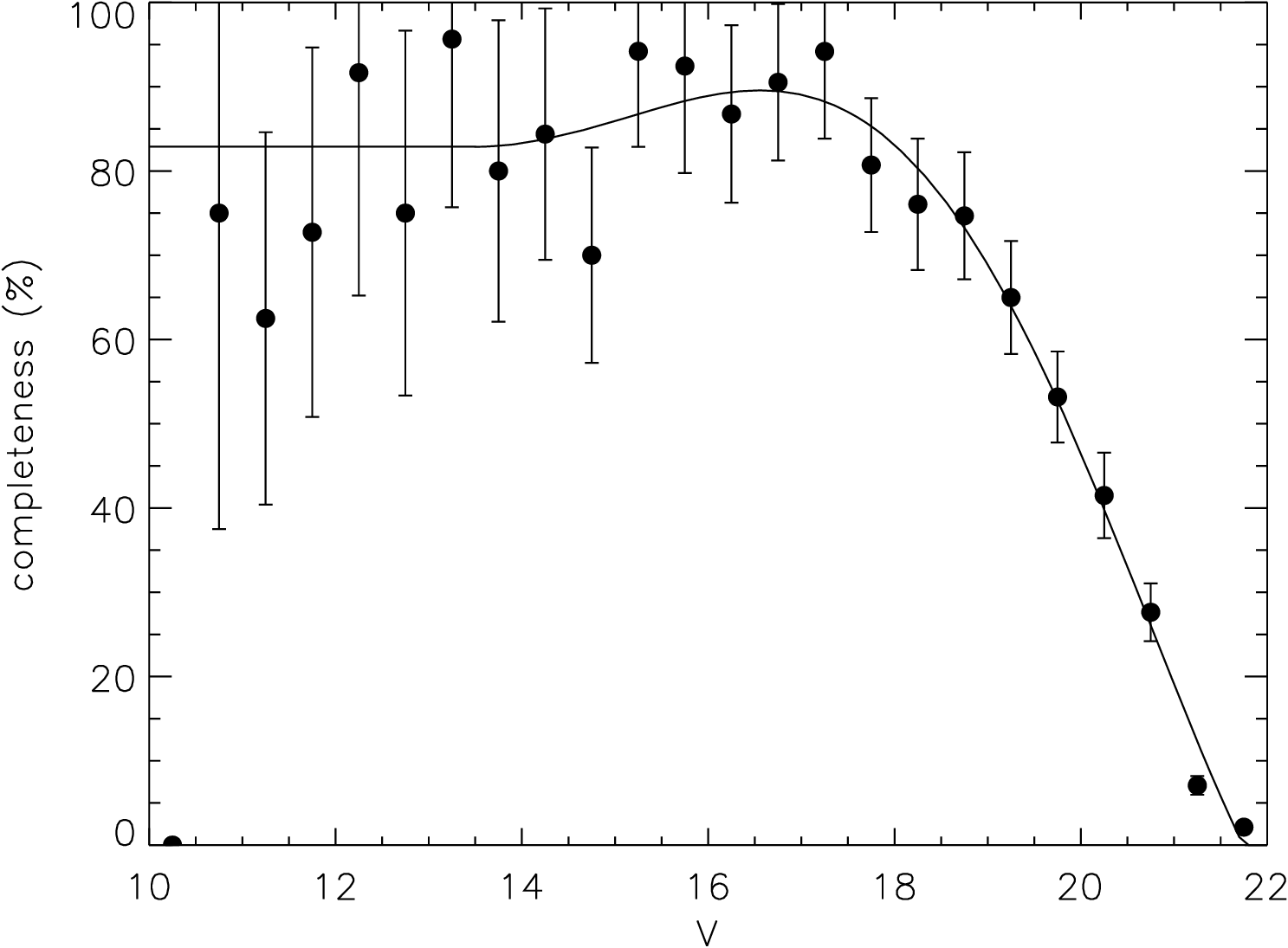}
\caption{The $V-$band completeness of the sub-sample of stars with \teff\ estimates (dots). The thick line is a 6th order polynomial fit for $V>13.5$, while the completeness has been assumed equal to a fraction $0.83$ for $V\leq14$.
    \label{fig:completeness_LAH}}
\end{figure}

In Figure \ref{fig:completeness_LAH} we show the HRD subsample completeness function. We fit a polynomial for $V>13.5$, while we assume the function to be constant, at 83\% completeness for $V\leq14$ given the high statistical uncertainties in the determination of the function.

We include 50\% of the sources at $V\simeq20$~mag and the completeness is zero for $V>22$~mag. The completeness is lower than 100\% for the brightest end of the sample, and this is due both to saturated sources (saturated in $I$ and not present in our catalog) or simply because of the lack of spectroscopy. The higher fraction at $15\lesssim V \lesssim18$~mag comes from the classification of M-type stars from the [TiO] photometric index, capable of detecting almost all the sources in this range and therefore providing a more complete spectral typing coverage than spectroscopy.

This completeness is used in the next sections, converted from being expressed in magnitudes to masses and ages, to correct the age distribution and the mass function.

\subsection{Spectral type-temperature relations}
\label{section:Luhman}
In Section \ref{subsection_spectraltypes} we have collected the spectral types for part of the sources present in our photometric catalog. These must be converted into \teff, because: 1) for deriving intrinsic colors and bolometric corrections (see Section \ref{section:bolometric_correction}) we rely on synthetic photometry on grids of stellar atmosphere models which are computed in terms of \teff; 2) also our model for the displacement of the colors caused by accretion luminosity -- which we use in the following sections to disentangle this effect for determining $A_V$ of the members -- is based on the same quantities.

\begin{figure}
\epsscale{1.1}
\plotone{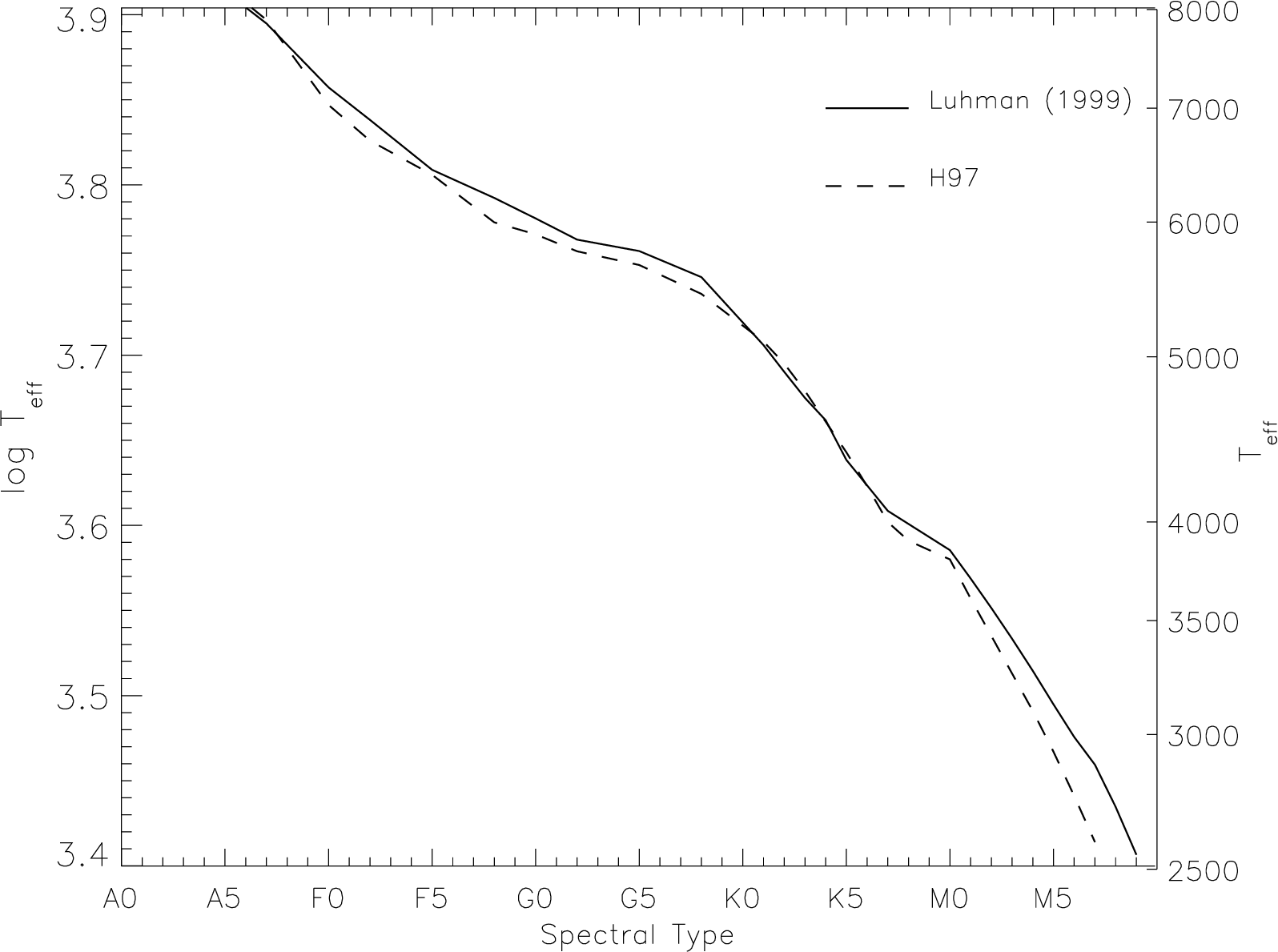}
\caption{Comparison between the spectral type-\teff\ relations used for this work \citep{luhman2003} and the one used on the ONC in H97 that was derived in \citet{cohencuhi79}.  The main differences rise with increasing spectral class, and can be several hundreds of Kelvin for late-M type stars. \label{fig:spt_t_rel}}
\end{figure}

To derive the effective temperatures from the available spectral types we use the spectral type vs. temperature relation of \citet{luhman2003}. This relation, qualitatively intermediate between giants and dwarfs for M-type stars, is derived empirically imposing that all the four members of the presumably coeval multiple system GG~Tau are located on the same BCAH98 \citep{BCAH98} isochrone \citep{luhman99}, and uses the relation from \citet{schmidt-kahler82} for types $<M0$ .
This is different from what used in H97, taken largely from \citet{cohencuhi79} (and supplemented with appropriate scales for O-type stars \citep{chlebowski1991} and late M-types \citep{kirkpatrick1993} and shifted in the K8-M1 range to be smoother). Figure \ref{fig:spt_t_rel} compares the spectral type - temperature relations of Luhman with the one of H97. The major differences are for the M spectral types, where Luhman predicts higher \teff.

Using the Luhman relation partially solves a problem encountered by H97, the systematic trend of the average extinction as a function of temperature with late type stars appearing bluer than the color predicted according to their spectra type for zero extinction. This effect in H97 was partially due to the temperature--intrinsic color transformation assumed in that paper, which we already improved as described in Section \ref{section:emprical_colors}; but only by using the Luhman relation between spectral types and \teff\ is it possible to remove the correlation between mean reddening and temperature, characterized by systematically lower -- even negative -- values of $A_V$ moving towards late-M spectral types.  The Luhman scale, predicting for a given late spectral type a higher  temperature and therefore bluer intrinsic color, reduces the number of sources whose observed colors cannot be matched by adding some reddening. A more detailed analysis on how the choice of the temperature scale affects the derived extinctions is presented in subsection \ref{section:comparison_av_results}.

\begin{figure*}
\epsscale{1.1}
\plotone{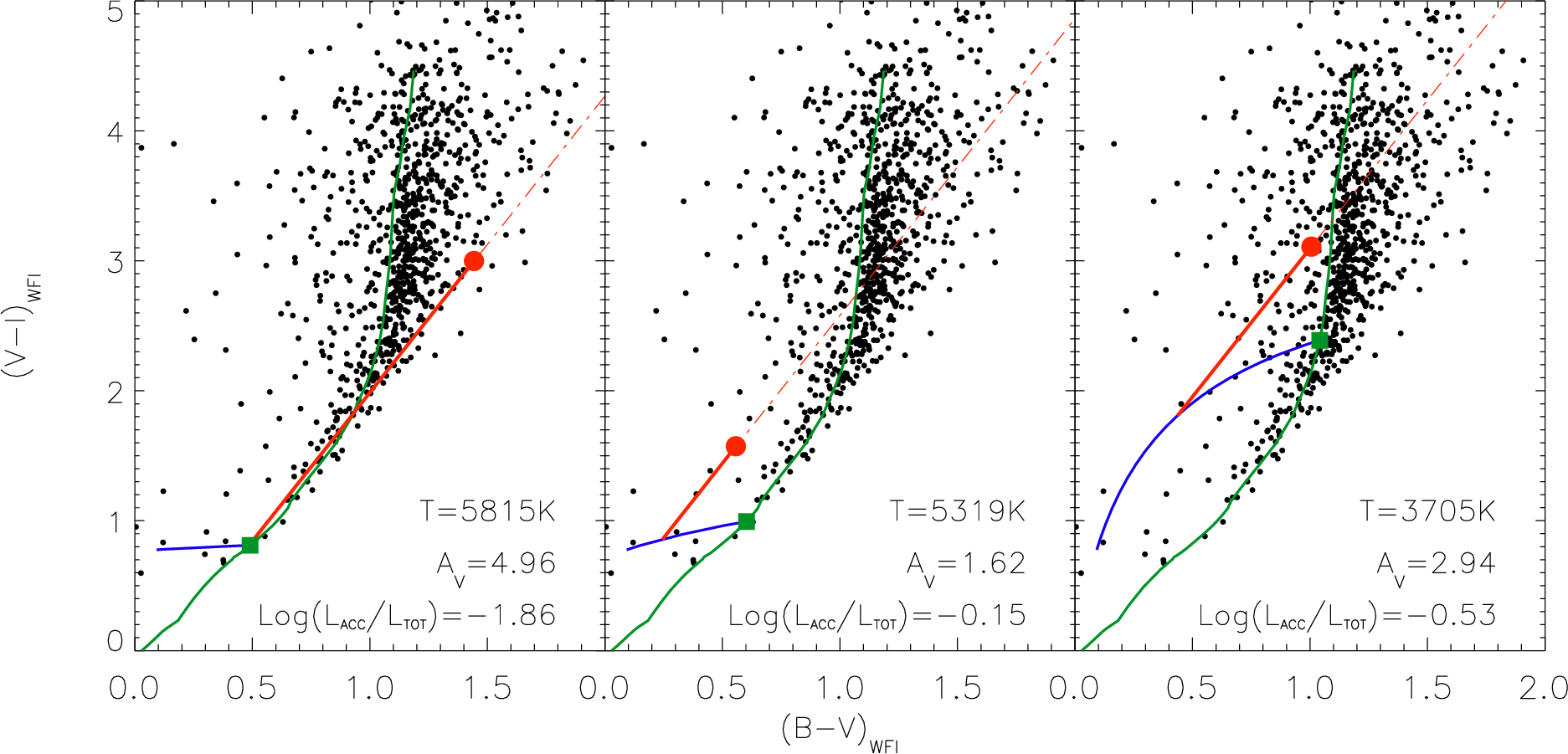}
\caption{Examples of the method applied for the determination of reddening taking into account accretion for three stars in our sample. The ONC sample with available \teff\ estimates (black dots) is plotted together with our reference model (green line). For each plot, the red circle indicates the observed colors of the considered star, and the green squares the predicted photospheric colors (for $A_V=0$ and $L_{\rm acc}=0$) given \teff\ of the star from its spectral type. The blue curve is the displacement, from the photospheric color, increasing $L_{\rm accr}/L_{\rm tot}$ from 0 to 1 (see Figure \ref{accretioncmd}). The intersection of the reddening direction (dash-dot line) applied to the star, and the accretion curve delimits the reddening vector (solid red line). In the three cases we report in the panels the known \teff\ and the derived extinction and accretion luminosity for the considered star.
\label{fig:evi_accretion_method}}
\end{figure*}

\subsection{Derivation of reddening and accretion}
\label{section:ReddAcc}

The reddening is defined as the difference between the observed color and the intrinsic one, for example the color excess $E(V-I) = (V-I)_{\rm obs}-(V-I)_{0}$ .
The intrinsic color $(V-I)_0$ is usually derived on the basis of  an assumed temperature-color relation.
In the presence of accretion, however, the source colors are also affected by $L_{\rm accr}$ as shown in Figure \ref{accretioncmd}, and since this effect makes the colors bluer, partially compensating the reddening caused by dust extinction, neglecting the accretion may lead to an underestimation of $A_V$.
Furthermore, in our analysis of the color-color diagrams we have accounted explicitly for surface-gravity effects and for accretion effects in all the photometric bands of our photometry. This enables us to assess both reddening and accretion more realistically than past methods, based on optical photometry, which assume that only one or the other of the effects dominates in the bands in which the observations have been obtained.

We have therefore developed a technique that consistently takes both $A_V$ and $L_{\rm accr}$ into account, providing at the same time an estimate of the accretion luminosity and reddening.
Given a star for which \teff\ is known, our  reference model gives its original position in the color-color diagram
for zero reddening and accretion; the measured location of the star in the color-color plots will be  in general displaced because of these effects. But under the assumption that they are uncorrelated of each other (the differential extinction depends mainly on the position of the star inside the nebula) and since we know how to model them (i.e., we know the reddening vector and we have a model for the accretion providing the displacement curve for all possible values of $L_{\rm accr}/L_{\rm tot}$), we can find the solution that produces the measured displacement with a unique combination of $A_V$ and $L_{\rm accr}/L_{\rm tot}$.

In practice, for each star in the color-color diagram, $(B-V)$ vs. $(V-I)$ we compute the track (like those shown in Figure \ref{accretioncmd}) representing the displacement computed for increasing $L_{\rm accr}/L_{\rm tot}$ from $0$ to $1$ for the exact \teff\ value of the star. Then we found the intersection between the reddening vector applied to the position of the star and the accretion curve. This provides the exact reddening vector, and its component along the $y$-axis gives the excess $E(V-I)$. The intersection point also gives a value of $L_{\rm accr}$ for the star, according to our model of the accretion spectrum. In this way, under the assumption that the accretion spectrum we have used is representative of all the stars and the reddening law is the same for all stars, the values of $L_{\rm accr}$ and $A_V$ for each star are uniquely derived.
Some examples of this method are plotted in Figure \ref{fig:evi_accretion_method}.

In a few cases this method cannot find a geometrical solution; this happens when the star is too blue or too red in $(B-V)$, so that the extinction line, applied backwards to the star, never crosses the curve computed for its temperature. Since the photometric errors are higher in $B$ than in $V$ or $I$, we attribute this effect to the inaccuracy of the $(B-V)$ color, and therefore we move the point along this axis until the closest intersection condition is found, respectively the tangency condition of reddening line and accretion curve, when the location of the star is too on the left. In the diagram, and the starting point along the accretion curve where $L_{\rm accr}=0$ when it is too on the right.

For some stars the extinction and accretion determination is degenerate. This is the case of Figure \ref{fig:evi_accretion_method}c, where the reddening vector intersect the accretion curve in 2 points, either low $A_V$ and low $L_{\rm acc}$ (the solution shown) or higher $A_V$ and $L_{\rm acc}$. In this cases, which involve only low-\teff\ stars, we always choose the first scenario.

In a few cases the derived extinction came out negative. This happens when the observed $(B-V)$ is higher and $(V-I)$ is lower than the locus for zero accretion at the star's temperature.
For 13\% of the sources the $B$ magnitude is not available and $A_V$ is therefore derived simply from the excess $E(V-I)$.

\begin{figure}
\epsscale{1.1}
\plotone{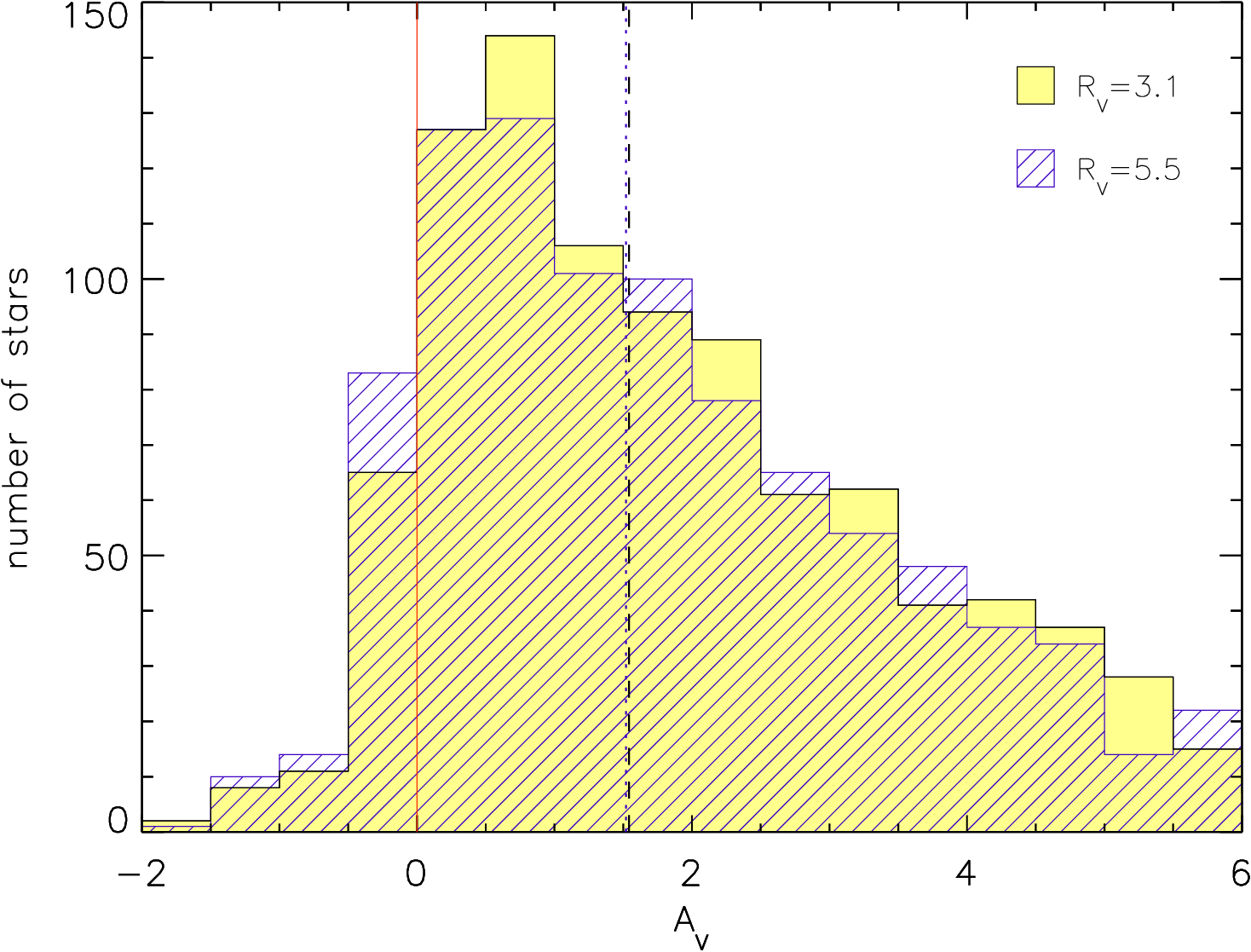}
\caption{Distribution of the extinction measured for the ONC stars, in terms of $A_I$ in the WFI photometric system, for values of $R_V=3.1$ and $5.5$. The dashed line shows the median value for the extinction for $R_V=3.1$, at $A_V=0.92$ mag; the dotted line is the median for $R_V=5.5$ at $A_V=0.97$ mag. \label{fig:reddeningdistribution}}
\end{figure}

In Figure \ref{fig:reddeningdistribution} the distribution of $A_V$ is shown, for the two choices of $R_V$. There is a small fraction of stars, as mentioned, for which we find a negative extinction. This, reasonably accounted for by the uncertainties, is however significantly smaller than in H97. In these cases we consider the star to have zero extinction for the derivation of the stellar parameters, as in H97.
The median of the distribution extinction for the whole sample is at $A_V\simeq1.5$~mag.

\begin{figure}
\epsscale{1.1}
\plotone{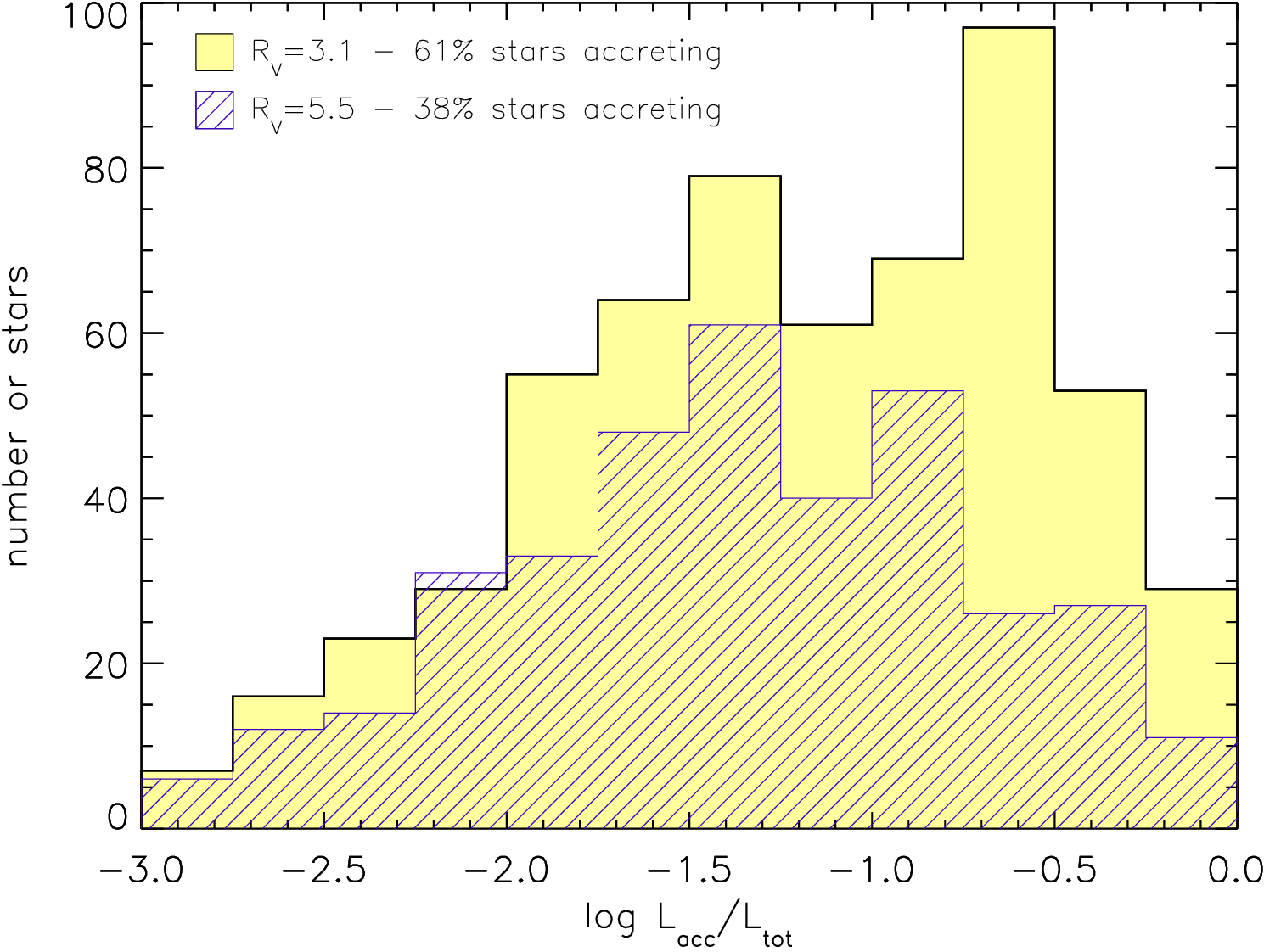}
\caption{Distribution of the accretion luminosities derived, for the two values of $R_V$.  \label{fig:histgram_Lacc}}
\end{figure}

In Figure \ref{fig:histgram_Lacc} the distributions of $\log L_{\rm accr}/L_{\rm tot}$ is shown, including all the stars for which the value of  $L_{\rm accr}$, derived from our geometrical method, is positive and big enough to lead to a displacement in the two color diagrams larger than the photometric error. We find about $60\%$ of the stars showing an excess attributed to accretion, with a relative fraction $L_{\rm accr}/L_{\rm tot}$ peaking at $5$--$25\%$. These values are comparable to those seen in the more rapidly accreting Taurus members \citep{herczeg2008}.  Increasing $R_V$ produces a decrease in both the fraction of accreting sources and the measured $L_{\rm accr}$, as a consequence of the steeper de-reddening vector in the $B,V,I$ plot.

We estimate the uncertainties in our determined extinction and accretion luminosity using again the geometrical method in the 2-color diagrams. Given the non-analytical nature of the method, we use a Monte Carlo (MC) approach to compute how the uncertainties in our data (both photometric errors and uncertainty in the spectral types) propagate to the determined values of $A_V$ and $L_{\rm accr}$. For each star we perturb randomly the photometry of each band with their photometric errors, assumed to be gaussian. We also consider the uncertainties in the spectral types from spectroscopy, of $\pm$1 sub-type, and we assign a temperature within this interval from a flat distribution. We iterate this approach 200 times for each star, deriving each time  the  $E(V-I)$ and $\log(L_{\rm accr}/L_{\rm tot})$ values. We then analyze the overall variation of the derived results. Our test shows that, on average, $\sigma_{E(V-I)}=0.12$ mag, with the uncertainty in the spectral typing being the dominant source of uncertainty. 
Concerning the accretion luminosity fraction $\log(L_{\rm accr}/L_{\rm tot})$, our MC test shows that the uncertainty can be up to 1 order of magnitude. We conclude that our method is fairly accurate in computing the extinction $A_V$, a most critical parameter for placing the members in the HRD, but provides only a rough  estimate of the accretion luminosities of the sources.
Other methods more directly based on accretion indicators (line emission, UV-excess,...) should therefore be used for an accurate analysis of the accretion activity of our sources.

\subsection{Systematic effects in deriving $A_V$}
\label{section:comparison_av_results}
\begin{figure*}
\epsscale{1.05}
\plotone{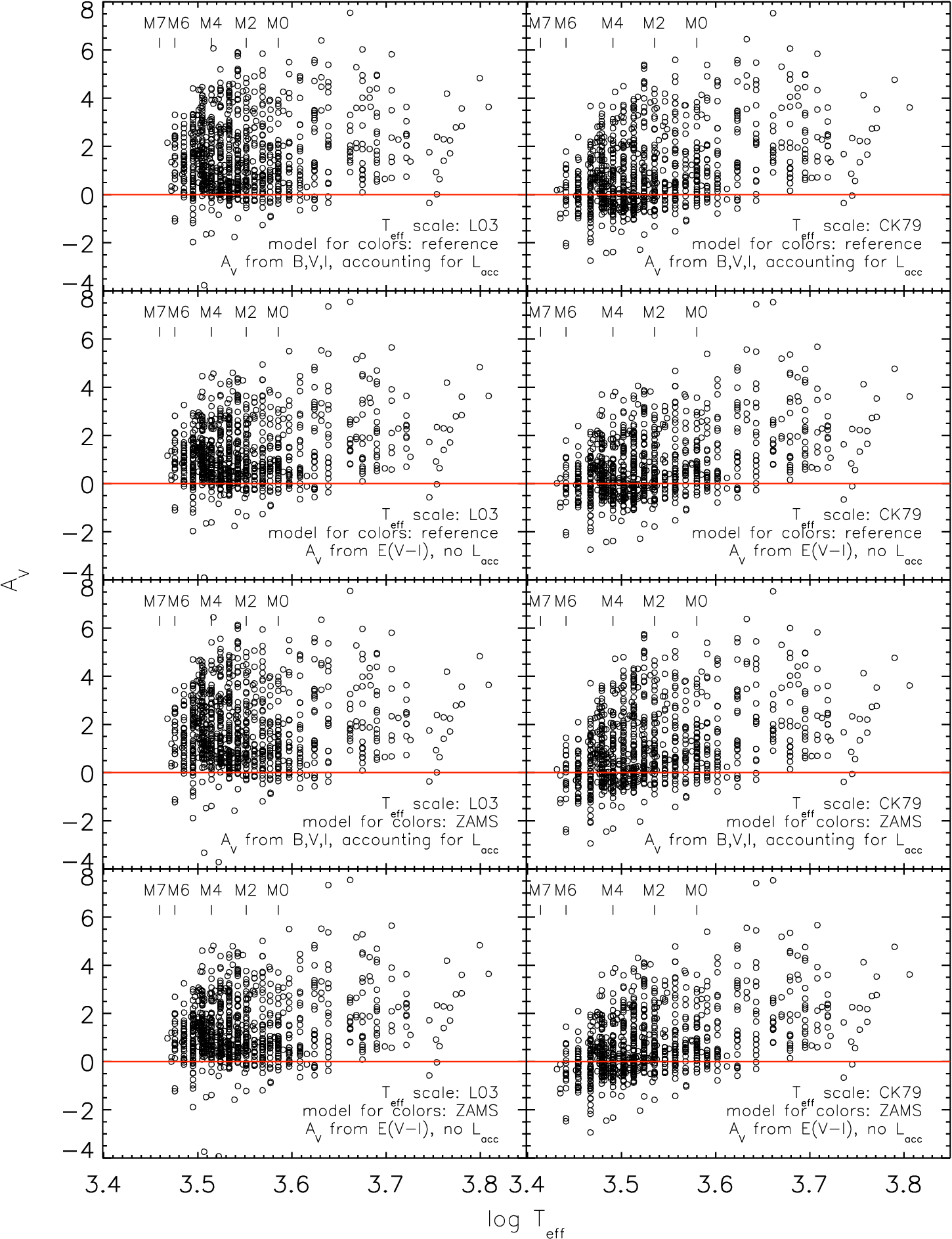}
\caption{$A_V$ vs. $\log$(\teff)\ plots for the ONC members, for all the combinations of changing:
\emph{a)} \teff vs. spectral type scale -- left panels using the Luhman scale, right panels using the Cohen \& Kuhi scale;
\emph{b)} intrinsic colors: -- rows 1 and 2 using our empirically calibrated reference model, rows 3 and 4 using the result of synthetic photometry constraining $\log g$ from the Siess ZAMS;
\emph{c)} method to derive $A_V$: - row 1 and 3 using our geometrical method to disentangle $A_V$ from $L_{\rm accr}$, row  2 and 4 neglecting $B$ magnitude and deriving $A_V$ from $E(V-I)$.
\label{fig:logt_evi_test}}
\end{figure*}

In the previous sections we have presented evidences that intrinsic colors valid for MS dwarfs are incompatible with the observed colors of the ONC, suggesting a gravity dependance of the optical color scale for M-type PMS stars. This is also predicted by atmosphere modeling, though with a limited accuracy. Thus we have calibrated the optical colors to best match our ONC data, defining our reference model for colors as a function of \teff.

We have presented a method to estimate $A_V$ from 3 photometric bands accounting for possible accretion luminosity, and we have mentioned that we chose a particular spectral type -- \teff\ relation to reduce the number of sources showing negative $A_V$. Here we analyze in more detail how all these aspects affect the derived reddening distribution.

We have considered all 8 possible combinations of either considering or not considering the 3 described improvements, which is: a) converting the spectral types into \teff\ using the \citet{luhman2003} scale or the \citet{cohencuhi79} one, used, e.g., in H97; b) using the intrinsic colors in $B$, $V$, $I$ band from our reference model or from the $\log g$ values of the ZAMS; c) deriving $A_V$ using the from $BVI$ disentangling accretion as described in Section \ref{section:ReddAcc} or simply neglecting the accretion and the $B$-band information and basing solely on the excess $E(V-I)$. In the case of the ZAMS model, we have first computed the displacements in the 2-color diagrams caused by accretion luminosity, in a similar way as for our reference model, shown in Figure \ref{accretioncmd}.

The results are presented in figure \ref{fig:logt_evi_test}, as a function if \teff\ to highlight systematic trends. For guidance, the top-left panel shows the results we use, and the bottom-right one reproduces qualitatively the findings using the same methods and relations as in H97, with a systematical lower $A_V$ towards the latest M-type stars.

It is evident that using the Luhman temperature scale for M-type stars instead of the Cohen \& Kuhi relation leads to systematically higher $A_V$, compensating very well the problem of deriving negative reddenings.

The differences we observed in the $A_V$ values changing the color scales from ZAMS to our empirical reference model are more modest, and Figure \ref{fig:logt_evi_test} alone does not allow to clearly justify the choice of one or another color scale. We remind the reader that the intrinsic colors from the ZAMS model do not reproduce the observed color-color diagrams, (see Figure \ref{fig:colorcolor}), and therefore can not be considered representative of the observed data. In particular, assuming the dwarfs colors leads to systematically higher accretion luminosities for cool members, predicting almost all M-type stars to have a positive value of $L_{\rm acc}/L_{\rm tot}$.

Finally, if we do not consider the color offsets due to on-going accretion (Figure  \ref{fig:logt_evi_test}, second and fourth row), the predicted extinctions are underestimated. This is clear from figure \ref{accretioncmd}: a positive value of $L_{\rm accr}$ leads to lower intrinsic colors (both $B-I$ and $V-I$), increasing the measured $A_V$.

\section{The H-R diagram of the ONC}
\label{sectionH-R}
The Hertzsprung-Russell diagram, representing the total luminosity emanating from the stellar photosphere versus the photospheric effective temperature, is the physical counterpart to the observational color-magnitude diagram. The transformation between the two requires the knowledge of a number of  quantities. In our case, having at our disposal \teff, we convert the observed magnitudes into total luminosities correcting for accretion and extinction (Section \ref{section:ReddAcc}), using our computed bolometric corrections for the WFI photometric system (Section \ref{section:bolometric_correction}), and considering the solar colors and bolometric magnitude. We therefore  compute the luminosity as follows:
\begin{eqnarray}
\label{eq:logl_loglsun}
\log \bigg(\frac{L}{L_\odot}\bigg)  = 0.4\cdot\big[M_{bol},_\odot - M_{bol},_\star\big] \\
 = 0.4\cdot\big[M_{I_{\rm WFI}},_\odot - M_{I_{\rm WFI}},_\star + BC_{I_{\rm WFI}},_\odot - BC_{I_{\rm WFI}},_\star\big] \nonumber \\
  = 0.4\cdot\big[M_V,_\odot - (V-I_{\rm WFI})_\odot - I_{\rm WFI} + \Delta I_{ACC} +  \nonumber  \\ A_{I_{\rm WFI}} + BC_{I_{\rm WFI}}(T) + DM\big] \nonumber
\end{eqnarray}
\noindent
where $I_{\rm WFI}$ and $V_{\rm WFI}$ are magnitudes referred to the  WFI system, while $V$ is $V_{\rm Johnson}$. The term $\Delta I_{\rm accr}$ (negative) is the $I$ band excess due to the accretion luminosity derived as shown in \S\ref{section:ReddAcc}, directly computed from synthetic photometry on the accretion spectrum normalized to the measured value of $L_{\rm acc}$ of each star, for the $2/3$ of the stars showing accretion in their colors. $\Delta I_{\rm accr}$ has a mean value of $-0.16 \pm 0.1$ mag, and is clearly zero for the rest of the sample.
\begin{figure*}
\epsscale{0.80}
\plotone{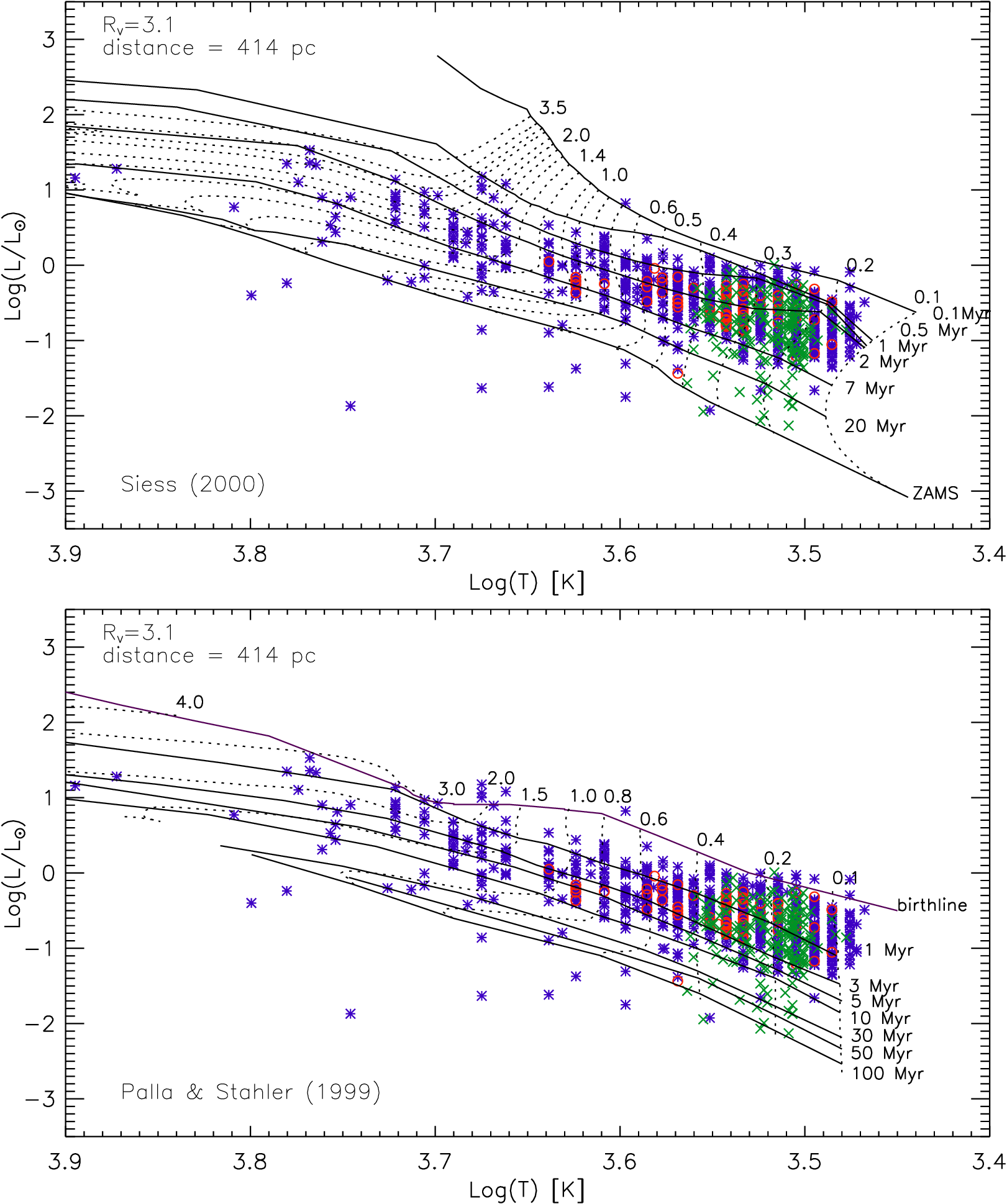}
\caption{Hertzsprung-Russel diagram for the ONC. Isochrones and tracks from \citet{siess2000} (upper panel) and \citet{pallastahler99} (lower panel) are overplotted, for masses from $0.1$M$_\odot$ to $3.5$M$_\odot$ and ages from $10^5$ years to the ZAMS in the Siess case, and from the birthline to $10^8$ Myr for the \citet{pallastahler99} models. Blue stars are spectral types from H97, excluding sources with a membership probability $<50\%$. Red circles indicate the additional sources we classify from our spectroscopic survey. Green crosses show the location of the M-type source, for which the temperature is determined from the [TiO] spectrophotometric index (see Paper~I). \label{fig:H-R}}
\end{figure*}
According to our convention for the bolometric correction, we have BC$_{I_{\rm WFI}},_\odot=0$. For the solar values we used $M_{V\odot} = M_{V{\rm (Johnson)}\odot} = 4.83$ \citep{galacticastronomy} and by means of synthetic photometry on our solar spectrum we find $(V_{\rm Johnson}-I_{\rm WFI})_\odot=0.80$.
Regarding the distance of the ONC, recent investigations based on accurate trigonometric parallax measurements \citep{hirota07,sandstrom07,menten2007} lead to a revision of the value  $d\simeq470$~pc adopted by H97. Amongst these, we used the most precise estimate  $d=414\pm7$ pc  \citep{menten2007}, based on a recent VLBA trigonometric parallax,  corresponding to a distance modulus DM$=8.085$.

As mentioned in Section \ref{section:subsampleanalysis}, we have excluded from our sample all the confirmed non-members, and we expect that the remaining contamination from foreground and background sources with unknown membership is relatively low ($\sim$2 -- 3\% of our sample).

The resulting HRDs, for the standard $R_V=3.1$ reddening law, are shown in Figure \ref{fig:H-R}. We mention that assuming the higher value of $R_V\simeq 5.5$ leads to predicted higher luminosities because, for a given color excess $E(V-I)$ the extinction $A_I$ is higher (see Table \ref{tablereddening}).
The upper and lower panels in Figure~\ref{fig:H-R} correspond to different PMS\ evolutionary models; top)\ the models by Siess, with ages from $10^5$ years to the ZAMS; bottom)\ the models of \citet{pallastahler99} (ages from $10^6$ to $10^8$ years), which include the deuterium birthline. Our sample reaches masses as low as $0.1{\rm M_\odot}$, extending to the intermediate-mass range up to several solar masses.

Figure \ref{fig:H-R} highlights the differences between the two families of evolutionary models in the comparison with our data. Whereas both are generally parallel to the median of the observed sequence, Siess predicts that isochrones of a given age are located at higher luminosities, which implies that the ages predicted for the cluster members are older. 
\citet{pallastahler99}, on the other hand, predict lower luminosities.

There are a few anomalous stars located in regions of the HRD not covered by the PMS isochrones. Stars located below the ZAMS can be background objects with no membership information. However, there are also some underluminous confirmed members. As explained, e.g., in \citet{kraus2009}, this effect seen in other PMS systems is due to the presence of dense circumstellar material (e.g., an edge-on circumstellar disks) which may both a) occult partially the photosphere; and b) scattering a significant amount of stellar flux - more efficiently for short optical wavelengths - along the line of sight. This hypothesis has received recent support the study of \citet{guarcello+2010}.  All the sources below the Siess ZAMS or the 100~Myr Palla isochrones are excluded from our further analysis.

We assess the overall uncertainty, from the observational errors, in the position of the stars in the HRD, using a  Monte Carlo approach similar to that adopted in Section \ref{section:ReddAcc}. We consider the uncertainty in the spectral type and the photometric errors, apply these effects randomly a number of times to each star, deriving the final position in the HRD as described above. For each source the methods produces a ``cloud" of points in the H-R diagram which represent the real scatter associated with the star.
\begin{figure}
\epsscale{1.1}
\plotone{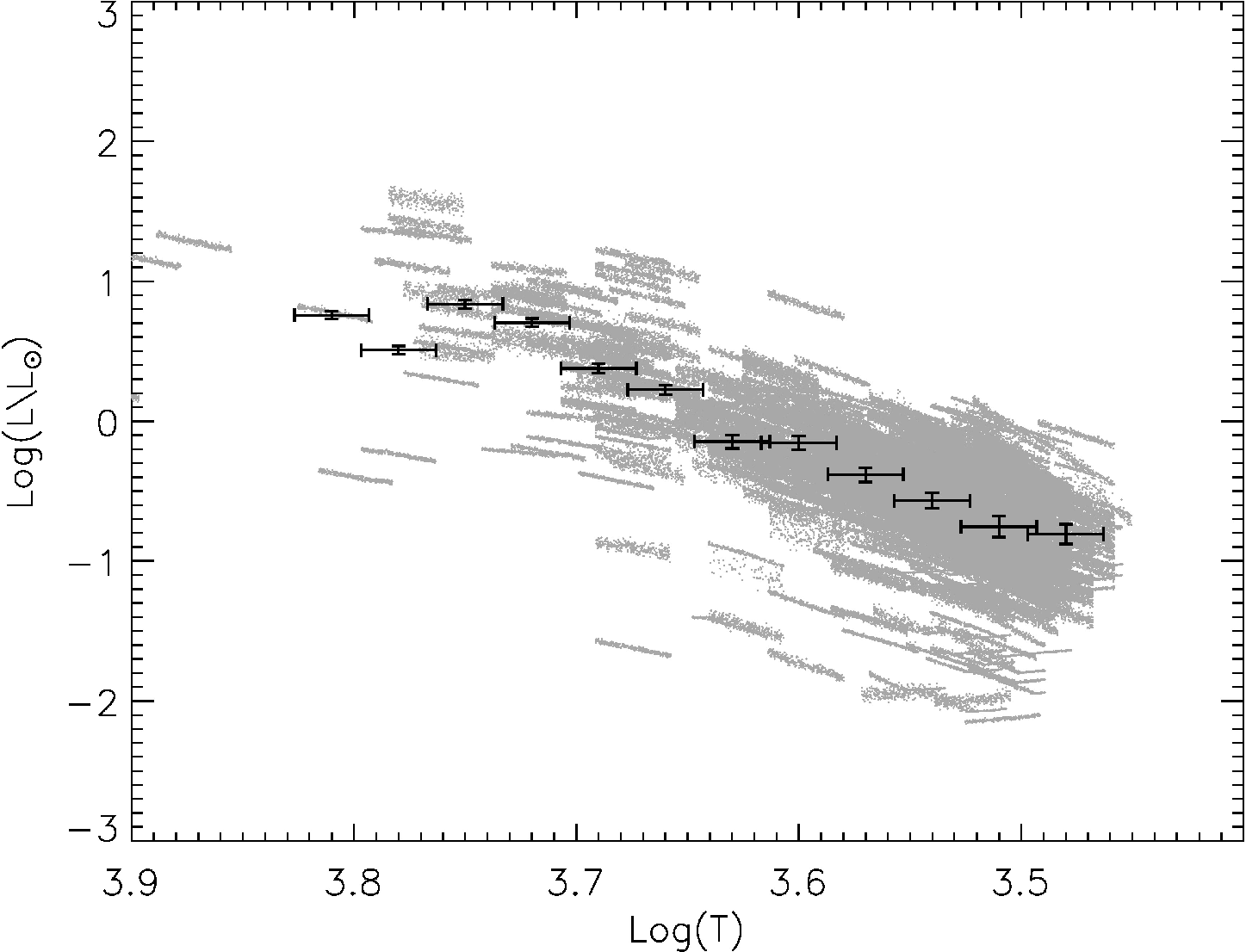}
\caption{Representation of the method we used for determining the average uncertainties that affects our determination of the H-R diagram. For each star we propagate the uncertainties simulating a ``cloud'' of determined positions in the diagram. The black points with error bars represent the average determined uncertainty in $\log L$ and $\log T$ in 12 bins of surface temperatures. \label{fig:mctest_H-R31}}
\end{figure}
Figure \ref{fig:mctest_H-R31} shows the result of an iteration. Each star is associated to a stripe of points, whose inclination and shape in general depend on the location of the star in the 2-color diagrams and its temperature. To understand better the general behavior of the resulting uncertainties, we divide stars in bins of $\log$(\teff) and compute the average spread in both \teff\ and $\log L$ with respect to the actual position of the stars of Figure \ref{fig:H-R}. This is represented with black points and error bars in Figure \ref{fig:mctest_H-R31}. It is evident that the uncertainty in \teff, which comes directly from the uncertainty in assigning spectral types, is usually much higher than the spread in $\log(L)$, even if this is influenced by the former (as mentioned in Section \ref{section:ReddAcc}, the errors in \teff\ usually contribute more than the photometric errors in the derivation of $A_I$, which affects $\log(L)$.

Concerning the luminosities, however, we must stress that there are several sources of uncertainties beyond the measurement errors, which involve both the data analysis and the physical nature of the members. In fact, our determination of $A_V$ and $L_{\rm accr}$ from the $BVI$ photometry relies on the assumption of a unique accretion spectrum in the modeling of the veiling. The SED of the accretion spectrum -- and therefore its broad band colors -- may actually be different from star to star. As mentioned in Section \ref{section:accretionCMD}, these variations should not affect much the extinction, and so $L_{\rm tot}$. We do not have at our disposal a representative sample of calibrated accretion spectra for the ONC, mostly due the bright nebular background against which a flux-calibrated excess would have to be measured relative to the stellar continuum. Therefore we are unable to characterize explicitly the error distribution on the luminosity caused by the accretion spectrum modeling.
Other sources of apparent luminosity spread are, moreover, circumstellar emission, scattered light, unresolved companions, variability and accretion history.
As a consequence, the errors on $L_{\rm tot}$ shown in Figure \ref{fig:mctest_H-R31} do not represent the maximum uncertainty, but the minimum error based on the uncertainties of our data that are known.
Only a quantitative knowledge of all the sources of luminosity spread could disentangle, therefore, apparent luminosity spread from true age spread. This problem, critical for the understanding of age distribution in the star formation history of young stellar systems \citep[e.g.,][]{hillenbrand2008,hillenbrand2009,baraffe2009} will not be discussed in detail in this work.

We can compare our derived luminosities with the values found by  H97, for the sources present in both catalogs.
\begin{figure}
\epsscale{1.1}
\plotone{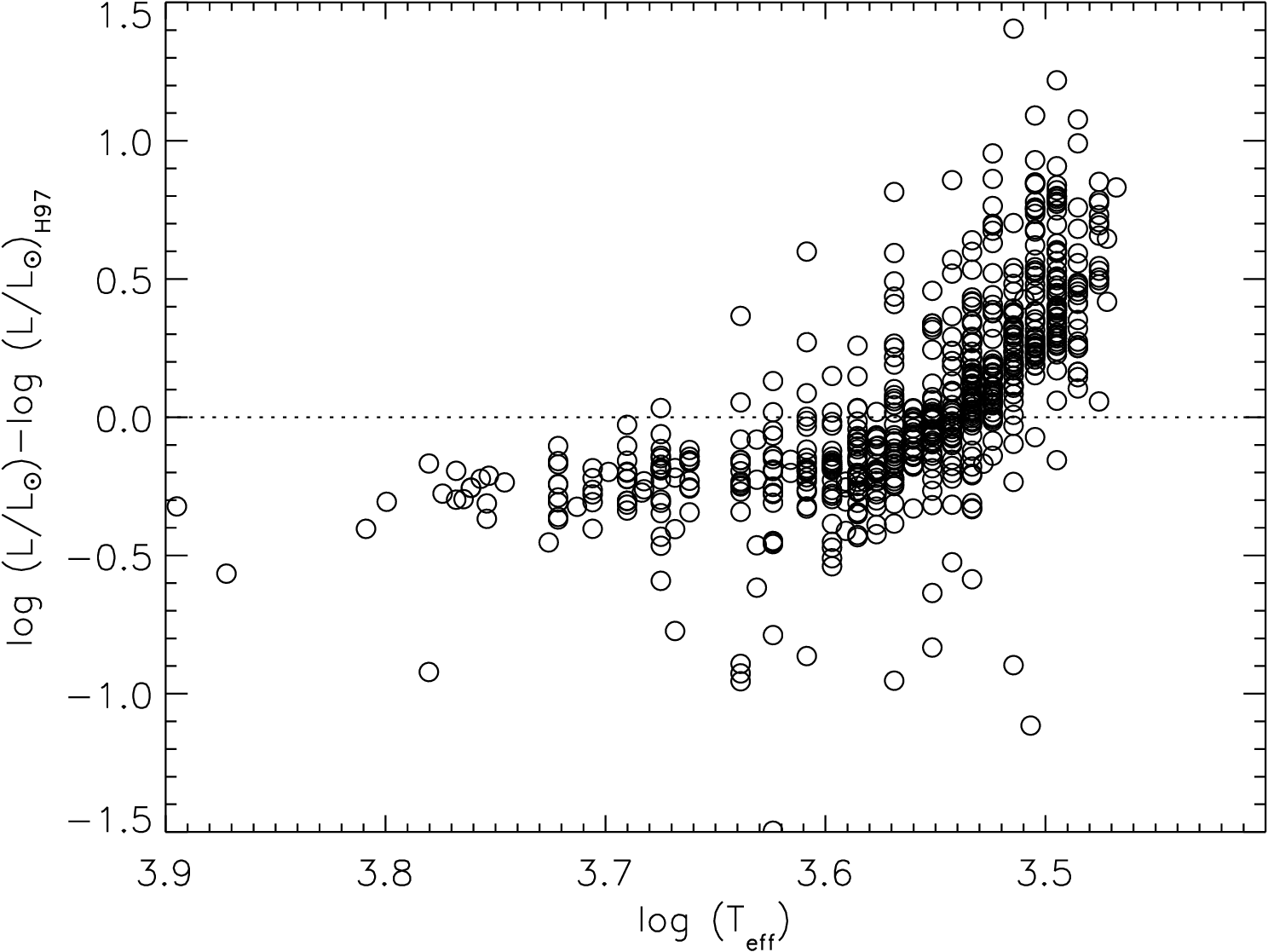}
\caption{Difference between bolometric luminosities derived from our calculations and from \citet{hillenbrand97}, as a function of \teff. The systematic trend is evident: for early type stars we derive a lower value of L, while for M-type stars our luminosities trend towards increasingly higher values towards lower temperatures, relative to H97. A number of
differences in the analyses contribute to this trend.  \label{fig:Lwfi_vs_Lh97.ps}}
\end{figure}
The results, as a function of \teff, are shown in Figure \ref{fig:Lwfi_vs_Lh97.ps}. Besides a modest scattering in the result, most probably due to stellar variability, a definite trend is evident: for $\log T_{\rm eff}>3.6$, corresponding roughly to K and earlier spectral types, we estimate a lower bolometric luminosity. This is in part caused by the lower distance we adopted for the ONC; however this accounts for only 0.11 dex of $\log L$. The changes we applied to the intrinsic colors and to the temperature scale do not affect critically this range of spectral types; therefore, this systematic effect is largely due to our bolometric corrections from $I-$band magnitudes. A different type of discrepancy appears for cold stars, where our analysis leads to higher luminosity than in H97. This is mostly due to  the higher extinctions compared to H97 that we measure in this range. Since the Luhman temperature scale we adopt predicts higher
\teff\ in this interval, our HRD is shifted both to higher $\log L$ and higher $\log T$ for M-type stars compared to H97, and continue to follow fairly well the isochronal slopes of evolutionary models.

\subsection{Completeness of the HRD}
\label{section:completeness_HRD}
\begin{figure}
\epsscale{1.1}
\plotone{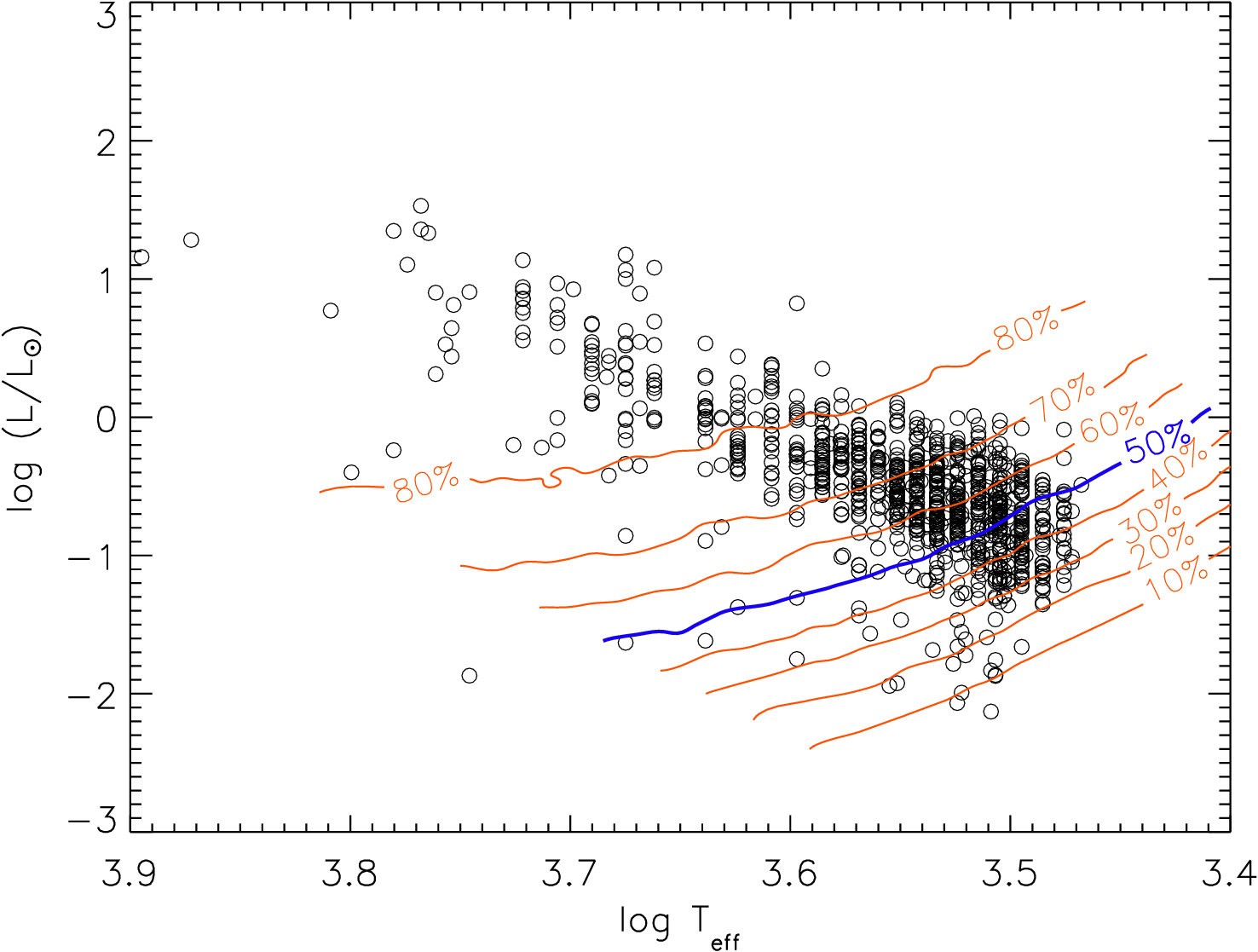}
\caption{Simulated completeness in the HRD  \label{fig:completeness_HRD}}
\end{figure}
In section \ref{section:completeness_LAH} the completeness of our spectro-photometric catalog is derived as a function of $V$ mag, i.e. for the CMD. This function, however, can not be directly translated in the HRD, because of differential extinction and non-linearity of intrinsic colors and bolometric corrections. Therefore we developed a statistical simulation that, for each point of the CMD, determines the probability for such star to be observed and present in our spectral type sample.

We proceed as follows. We consider a uniform, dense, cartesian grid of points in the HRD. For each of these points we find its counterpart in the CMD for $A_V=0$, converting \teff\ and $\log L$ into $(V-I)_0$ and $I_0$ using the intrinsic colors of our reference model and applying backwards the bolometric correction and the distance modulus. In reality, a star having the considered \teff\ and $\log L$ of the grid point is not necessarily observed in this position of the CMD, but reddened along the extinction direction of an amount that follows the distribution of $A_V$ inside the nebula. We consider as a representative distribution of $A_V$ the distribution we determined in section \ref{section:ReddAcc} limited for the brightest half (in $\log L$) of the population. This is because the reddening for intrinsically faint stars in our sample is biased towards low values, since such sources with high $A_V$ fall below our detection limit. For each point in the transformed grid in the CMD we utilize a Monte Carlo approach for applying the extinction, creating 1000 simulated reddened stars, with values of $A_V$ drawn from this distribution. Therefore, this set of aligned, simulated stars correspond to the all the positions in the CMD where a star with the considered \teff\ and $\log L$ can be located, following the true extinction distribution. We assign the completeness computed in Section \ref{section:completeness_LAH} to all the 1000 simulated stars according to their position in the CMD. The average of the 1000 values of completeness correspond statistically to the real probability, for a source with the considered \teff\ and $\log L$, to be detected in our catalog and assigned a spectral type. Iterating the same method for all the points in the HRD, we populate a 2D distribution of completeness in this plane.

The result is shown in Figure \ref{fig:completeness_HRD}. Our results show that the completeness decreases towards both low $L/L_{\odot}$ and low \teff.

\subsection{Distribution in the mass-age plane}
Using the models of \citet{siess2000} and \citet{pallastahler99}, we assign an age and mass to each star. To this purpose, we compute 2D interpolated surfaces with age and mass given in function of $ \log (T_{\rm eff}), \log (L/L_\odot).$
In Table~\ref{table:hrresults} we present our results for the stars of our sample.

\begin{sidewaystable}
\caption{Positions in the H-R diagram and values of mass and age form Siess (2000) and Palla \& Stahler (1999)}
\begin{tabular}{rrrrrrr|rr|rr|rrrl}
\multicolumn{7}{c}{ } & \multicolumn{2}{|c|}{Siess} & \multicolumn{2}{c|}{Palla} & \multicolumn{4}{c}{ }  \\
\multicolumn{1}{c}{ID} & \multicolumn{1}{c}{RA} & \multicolumn{1}{c}{Dec} & \multicolumn{1}{c}{$\log T_{\rm eff}$} & \multicolumn{1}{c}{$\log L$} & \multicolumn{1}{c}{$A_V$} & \multicolumn{1}{c}{$\log \frac{L_{\rm acc}}{L_{\rm tot}}$} & \multicolumn{1}{|c}{M} & \multicolumn{1}{c}{$\log$~age} & \multicolumn{1}{|c}{M} & \multicolumn{1}{c|}{$\log$~age} & \multicolumn{1}{c}{Spectral Type} & \multicolumn{1}{c}{H97 ID} & \multicolumn{1}{c}{mem  \tablenotemark{2}} & \multicolumn{1}{c}{Spectral Type} \\
\multicolumn{1}{c}{} & \multicolumn{1}{c}{J2000} & \multicolumn{1}{c}{J2000} & \multicolumn{1}{c}{K} & \multicolumn{1}{c}{$L_\odot$} & \multicolumn{1}{c}{mag} & \multicolumn{1}{c}{ } & \multicolumn{1}{|c}{${\rm M_\odot}$} & \multicolumn{1}{c}{yr} & \multicolumn{1}{|c}{${\rm M_\odot}$} & \multicolumn{1}{c|}{yr} & \multicolumn{1}{c}{from:  \tablenotemark{1}} & \multicolumn{1}{c}{} & \multicolumn{1}{c}{\%}  & \multicolumn{1}{c}{} \\
\hline
    1 &  5~35~47.02 & -5~17~56.91 &    3.615 &    2.739 &    7.263 &      $<-5$ &      ... &      ... &      ... &      ... &        H97 &      992 &        0 & K3-M0I ...\\ 
    2 &  5~35~20.71 & -5~21~44.45 &    4.251 &    2.628 &    2.184 &      -1.31 &      ... &      ... &      ... &      ... &        H97 &      660 &       99 & B3 ...\\ 
    3 &  5~35~05.21 & -5~14~50.37 &    3.768 &    1.529 &    2.665 &      -0.53 &    2.918 &    6.326 &    2.850 &    5.854 &        H97 &      260 &       97 & F8 ...\\ 
    5 &  5~35~21.32 & -5~12~12.74 &    3.768 &    1.359 &    2.027 &      -0.79 &    2.570 &    6.469 &    2.450 &    6.226 &        H97 &      670 &       70 & G8: ...\\ 
    8 &  5~34~49.98 & -5~18~44.61 &    3.936 &    1.625 &    2.511 &      -0.72 &    2.342 &    6.684 &    2.400 &    6.434 &        H97 &      108 &       92 & A3 ...\\ 
   10 &  5~34~39.76 & -5~24~25.66 &    3.662 &    1.045 &    0.898 &      -1.57 &    1.930 &    5.669 &      ... &      ... &        H97 &       45 &        0 & K4 ...\\ 
   11 &  5~35~16.97 & -5~21~45.42 &    3.969 &    1.371 &    1.345 &      -0.72 &    2.155 &    6.941 &    2.250 &    6.695 &        H97 &      531 &       99 & A0 ...\\ 
   15 &  5~35~05.64 & -5~25~19.45 &    3.699 &    0.926 &    0.769 &      -1.49 &    2.515 &    6.209 &    2.487 &    5.667 &        H97 &      273 &       99 & K0 ...\\ 
   16 &  5~35~20.21 & -5~20~57.09 &    3.780 &    1.348 &    4.239 &      -0.22 &    2.432 &    6.551 &    2.400 &    6.285 &        H97 &      640 &       99 & K0 ...\\ 
   \nodata & \nodata & \nodata & \nodata & \nodata & \nodata & \nodata & \nodata & \nodata & \nodata & \nodata & \nodata & \nodata & \nodata
\end{tabular}
\tablecomments{Only results assuming $R_V=3.1$ are included. (This table is available in its entirety in a machine-readable form in the online journal. A portion is shown here for guidance regarding its form and content.)}
\tablenotetext{1}{H97 refers to spectral types from Hillenbrand (1997), NEW to stars classified from our spectroscopy, TiO refers to M-type stars classified from the [TiO] photometric index.}
\tablenotetext{2}{Membership probability, as in H97. }
\label{table:hrresults}
\end{sidewaystable}

\begin{figure*}
\epsscale{0.9}
\plottwo{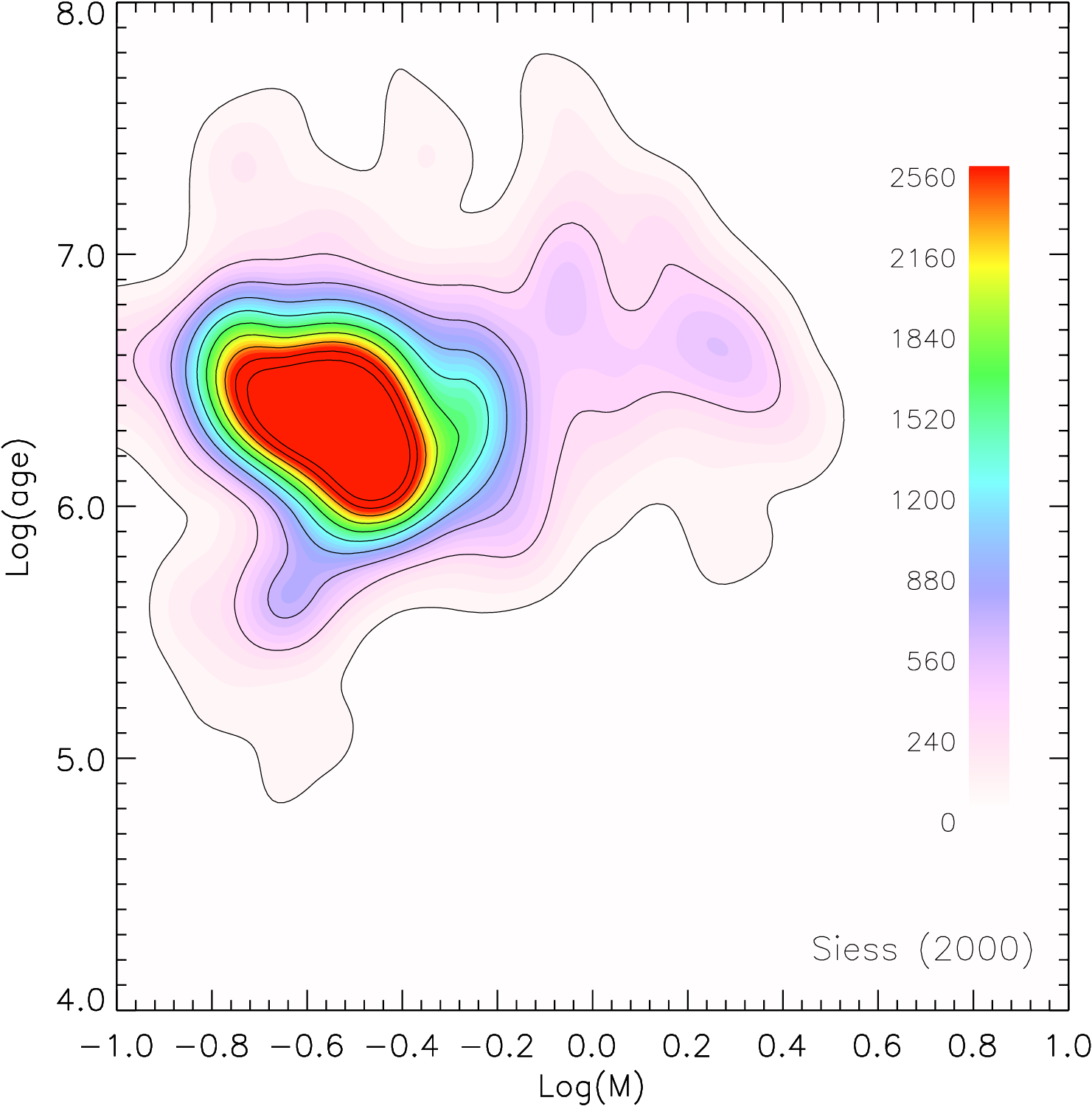}{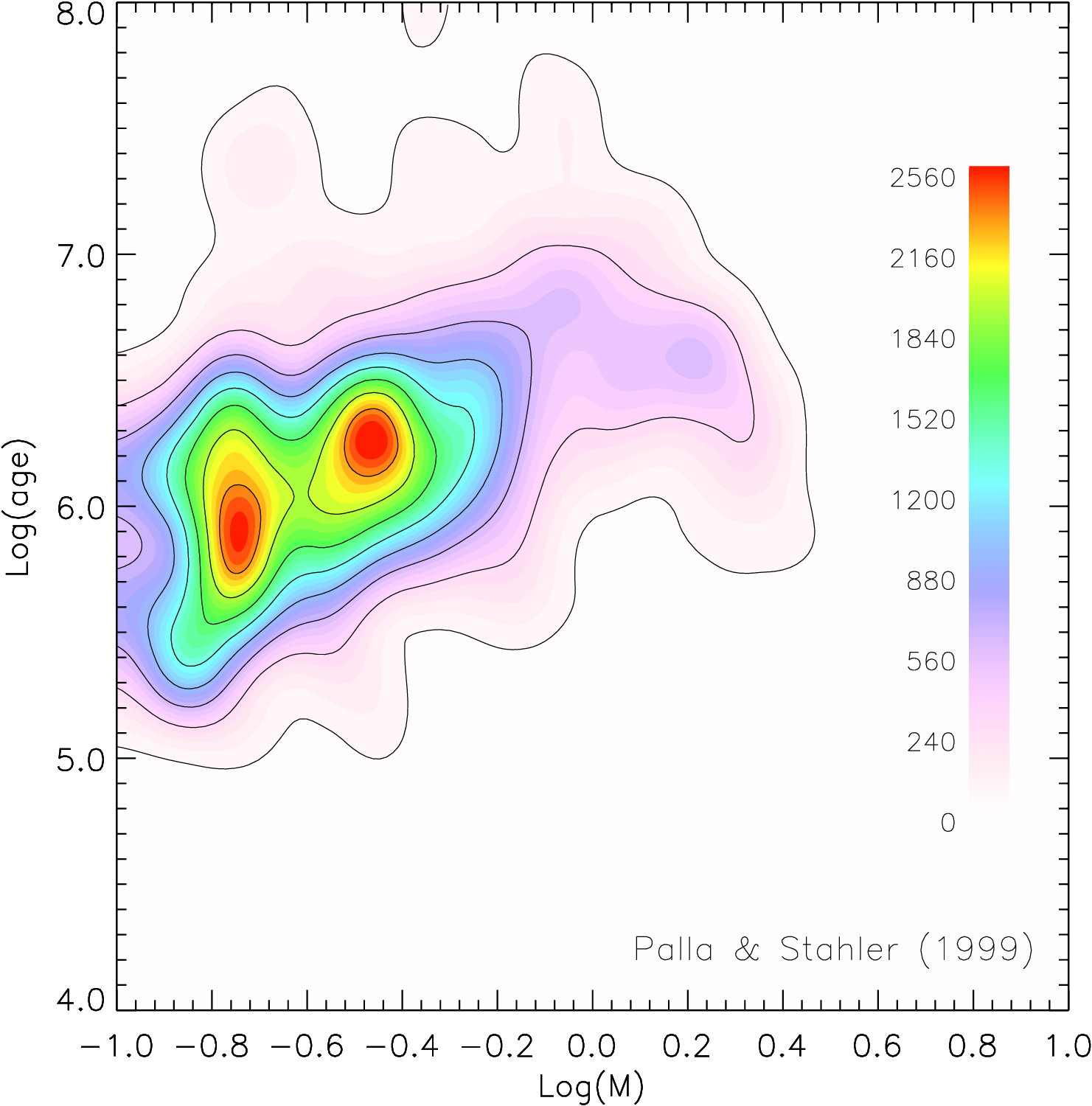}\\
\plottwo{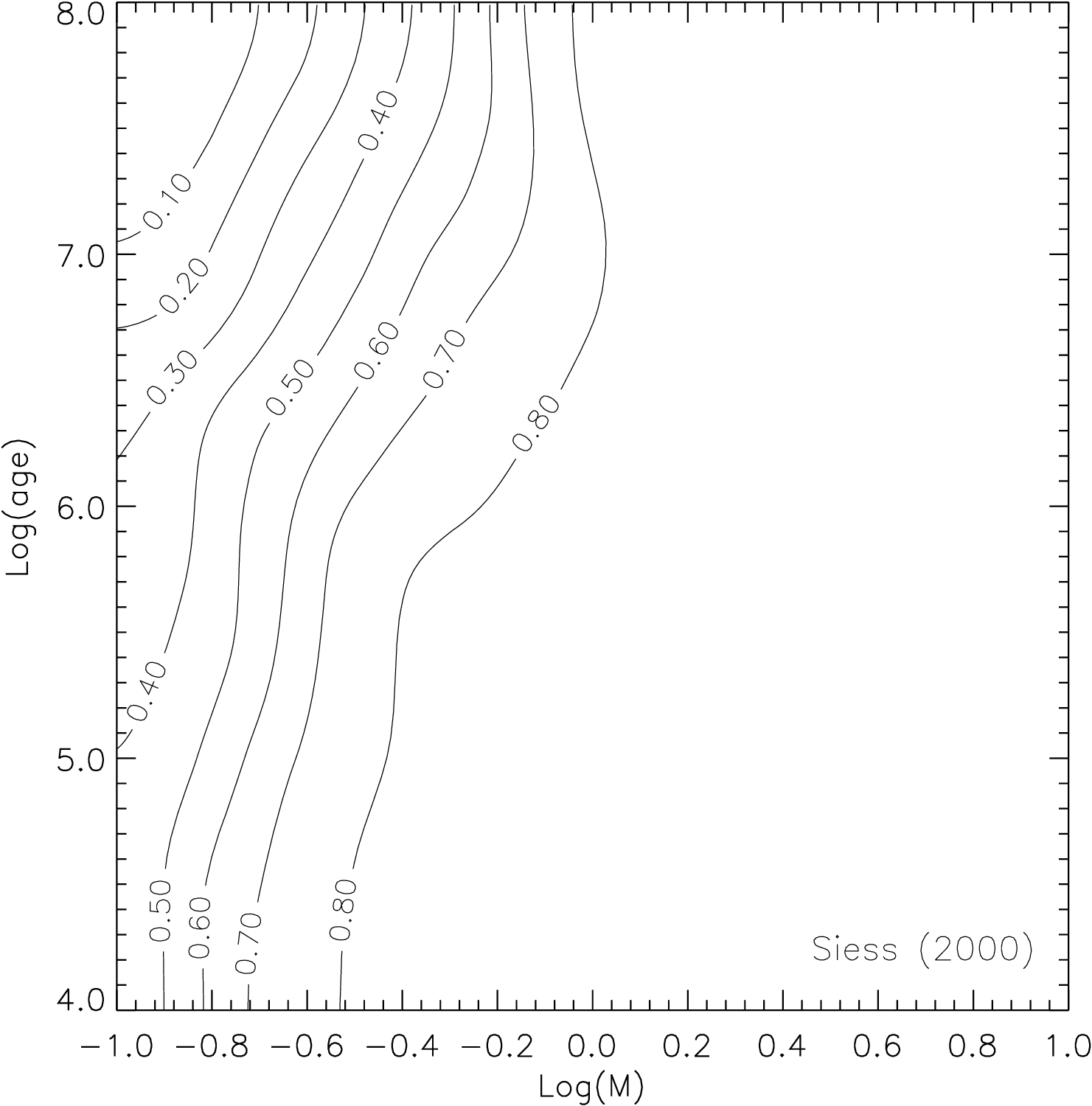}{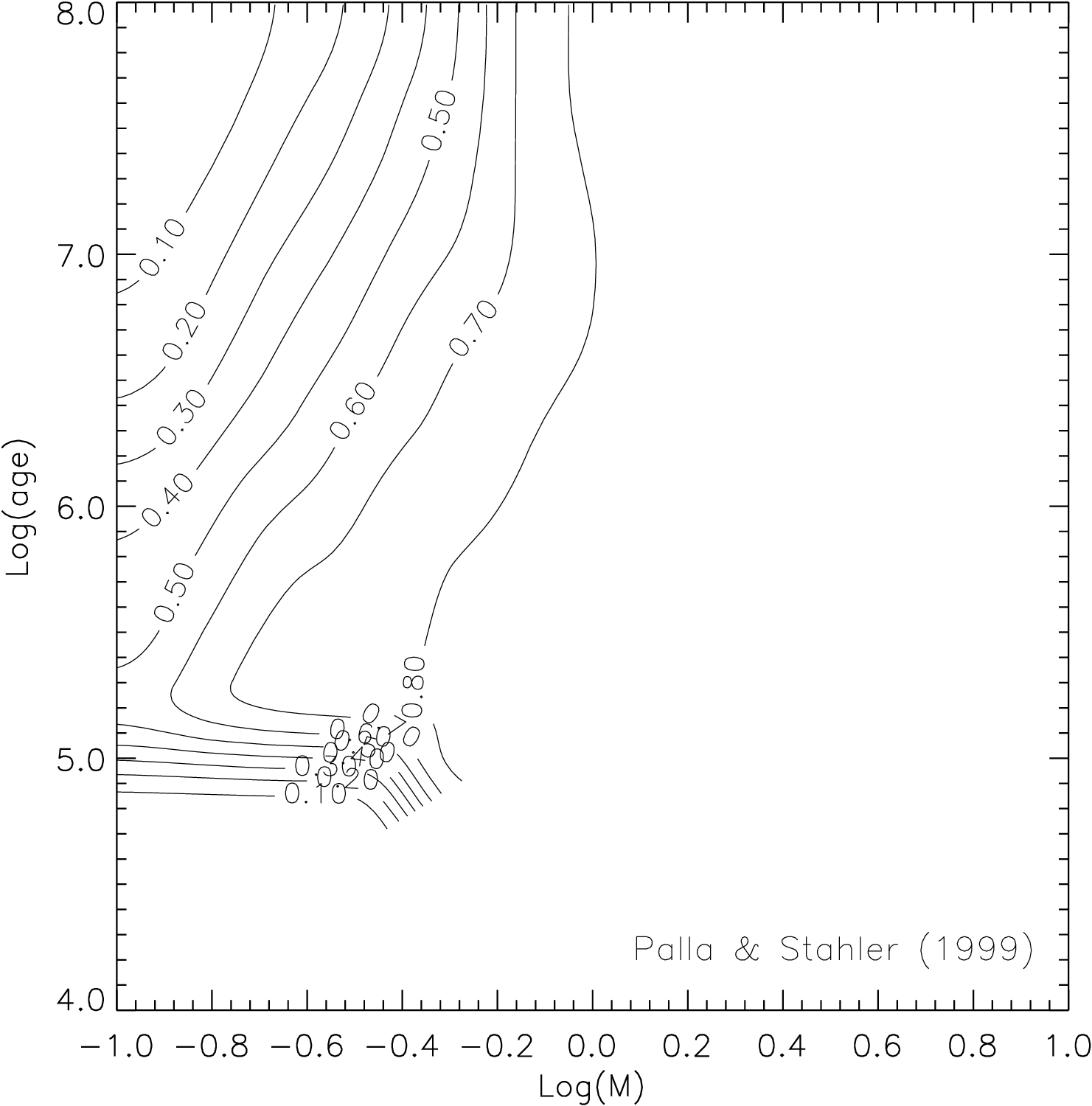}\\
\plottwo{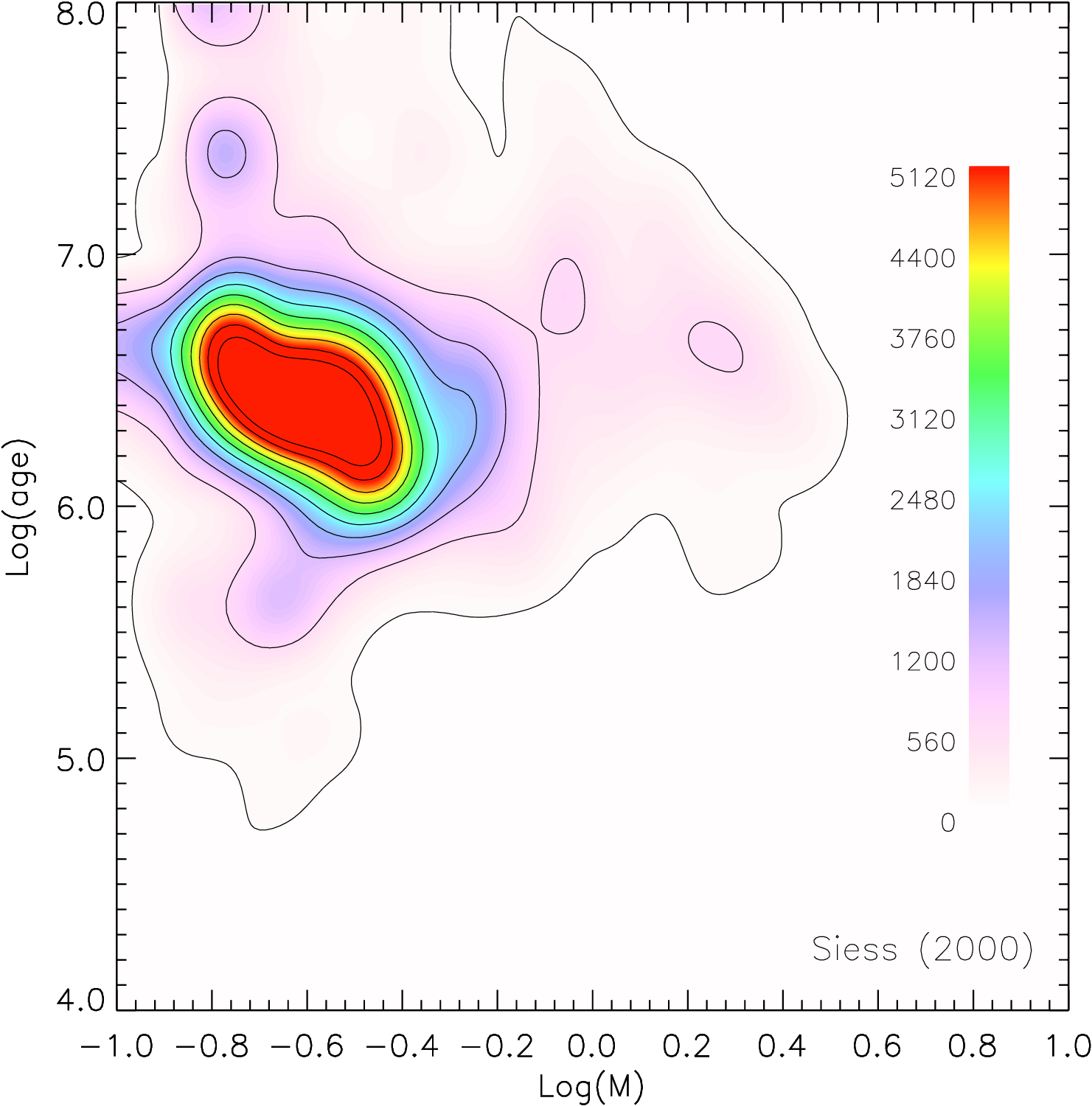}{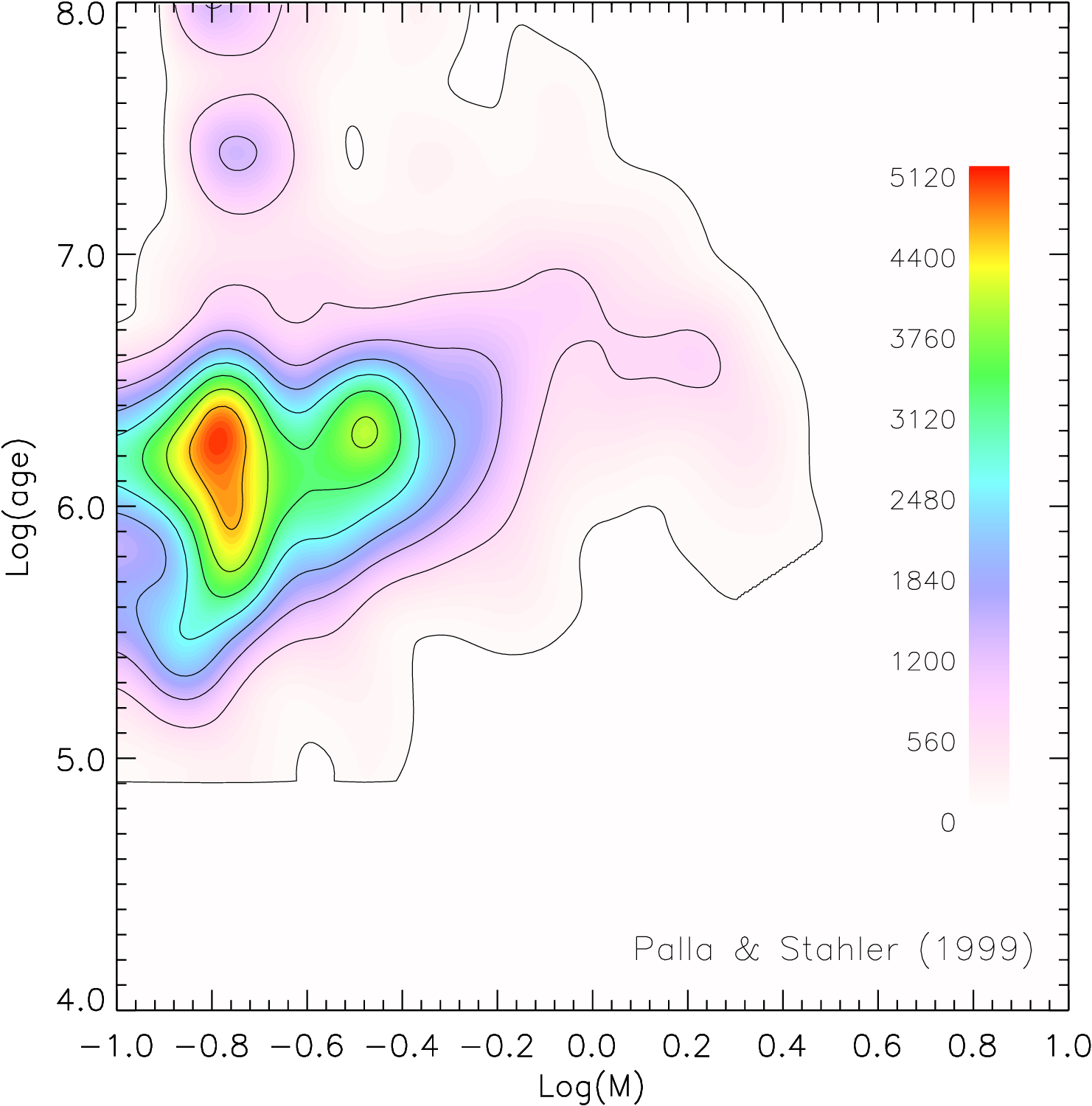}
\caption{\footnotesize  {\em Upper panels:}  density map in the plane $\log (age)$ vs. $\log (mass)$ according to our results using \citet{siess2000} and \citet{pallastahler99} evolutionary models. Stars are counted into a grid of uniform bins spaced $0.1$ in $\log (M)$ and $0.25$ in $\log (age)$, and the derived map has been oversampled, interpolated and smoothed with a kernel sized $0.13\times0.32$. The color scale reported is in units of number of stars per unit of $\log (M)$ and per unit of $\log (age)$. It is evident that in the Siess case the age distribution is uniform (besides a broad age spread at all the masses), but for the Palla \& Stahler models there is an evident correlation between age and mass, predicting younger ages for very low mass stars and older for intermediate masses.
{\em Middle panels:} contour plots of the completeness function in the mass-age plane, computed for the evolutionary models of \citet{siess2000} and \citet{pallastahler99}, and for $R_V=3.1$, from our simulation.
The apparent vanishing completeness in the bottom-left corner of the Palla \& Stahler case is due to the lack of information of these evolutionary tracks for young ages and very low masses, but this does not affect our results given that our sample includes no stars in this range.
{\em Bottom panels:} the density maps of the upper panels normalized dividing it by the completeness. \normalsize \label{fig:mass_age_relation}  }
\end{figure*}

We analyze our masses and ages, and study their distribution in the mass-age diagram; this allows us also to investigate possible correlations between the two quantities.

For this purpose, we produce a number density map in a $\log (age)$ vs. $\log (M)$ plot, counting the stars present into uniform bins of width 0.25 dex in $\log (age)$ and 0.1 dex in $\log (M)$. A boxcar smoothing with a kernel size of $1.3$ times the original grid size is applied to smooth out local variations due to statistical noise.

The result is shown in Figure \ref{fig:mass_age_relation}, upper panels, with density contours highlighted.
The outermost solid contour corresponds to 1 star per bin, the second 5, and the third 9, corresponding to poisson uncertainty of $100\%$, $45\%$, and $33\%$. Evidence of correlations should be investigated in the inner part of the map, where the majority of the objects are positioned.
Here it is evident that whereas Siess models do not show a global trend, in the Palla \& Stahler case the low-mass stars turn out to be systematically younger than the intermediate masses, with an average age changing by more that one dex along our mass range.
H97 reported a similar inconsistency in their analysis, based on the evolutionary models of \citet{dantona-mazzitelli94}.
The fact that the age-mass correlation is found using Palla \& Stahler models while not seen in our results from Siess isochrones could imply that the latter should be considered in better agreement with our data.

However, as discussed in \citet{pallastahler99}, the completeness of the sample can potentially bias the mass-age relation. This is especially true in the very-low mass regime, as seen in Figure \ref{fig:completeness_HRD}, as older PMS stars have lower luminosity than younger objects and therefore are more likely to be excluded from a stellar sample due to sensitivity limits. We therefore investigate if the incompleteness of our stellar sample could be responsible for the observed mass-age relation in order to correct the derived results.

We compute the completeness in the mass-age plane following an analogous method as for the HRD (Section \ref{section:completeness_HRD}), but considering instead of a uniform grid of $\log$\teff\ and $\log L$ values, a cartesian grid of masses and ages in the HRD, identical to the one used for Figure \ref{fig:mass_age_relation}a,b. The result is now model-dependent, since a given mass-age point is located in different positions of the HRD according to the two sets of evolutionary models.

The result is shown in Figure \ref{fig:mass_age_relation}, middle panels, interpolated and smoothed in the same way as for the density maps.
Results confirm that completeness decreases toward lower masses and higher ages, in a qualitatively similar fashion to the observed mass-age correlation found from our data with Palla \& Stahler isochrones. Differences between the results obtained with the two sets of evolutionary models are due to the different shape of tracks and isochrones, especially in the low-mass range (see the HRDs of Figure \ref{fig:H-R}). The cutoff for the youngest ages in the completeness derived using Palla \& Stahler models is due to the lack of the early PMS phases in  this family of models.

The ``normalized" mass-age map, computed by dividing our preliminary results of Figure \ref{fig:mass_age_relation}a,b by the completeness map, is shown in the bottom panels of the same Figure. While the shape of the distributions in the mass-age plane and the average age for the ONC do not change by a large quantity, due to the smooth variation of completeness against mass and age, we find that the correlation between mass and age for the Palla \& Stahler models reduces significantly, and below $1 {\rm M_\odot}$ the trend of age with mass is less evident, while, on the contrary, for the Siess models the completeness correction suggests an anti-correlation between ages and masses, hinting for higher predicted ages for low-mass stars.

In the high-mass end of these plots ($M \gtrsim 1$M$_\odot$) there is still a tendency toward a higher measured average age, but we do not consider this as a solid evidence for a measured mass-age correlation. This for two main reasons: on the one hand, intermediate stars of young ages are not covered by the evolutionary models of  Palla \& Stahler, and on the other hand for  low-mass and high-age fraction of the population the completeness we determined is too low to be correctable using our method. In conclusion, we do not find evidences of an age trend with respect to stellar mass in the ONC, both using Siess and Palla \& Stahler evolutionary models.

The apparent over-density evident in the completeness-corrected map at $\log M=0.8M_{\odot}$ for $\log({\rm age})>7$ is not relevant, being solely due to statistical noise.

\subsection{The age distribution}
\label{section:age_distribution}
\begin{figure}
\epsscale{1.1}
\plotone{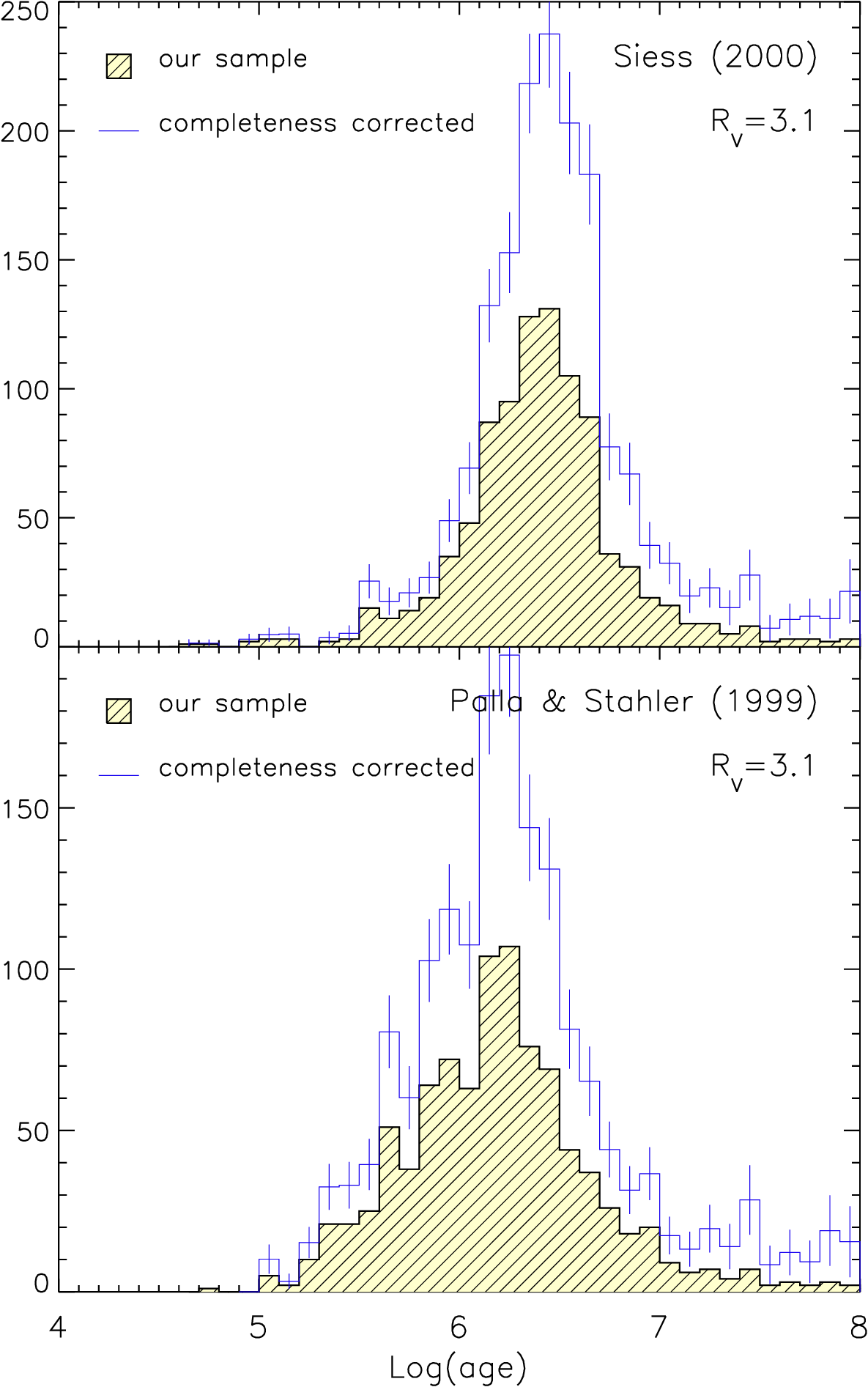}
\caption{Age distribution for the ONC population. \emph{Top panel:} ages derived using the \citet{siess2000} evolutionary models; \emph{bottom panel:} the same using \citet{pallastahler99}. The filled histograms show the distributions of ages in our sample. The open histograms are the same after correcting for incompleteness, and the statistical uncertainties are shown for each age bin.    \label{fig:histogram-age}}
\end{figure}

In Figure \ref{fig:histogram-age} we present the measured age distribution and the same after correcting for the completeness $C(\tau)$, for the two families of PMS models, as we now describe.
From the 2D completeness functions in the mass-age plane, we can now derive the overall completeness in ages.
If $C(M,\tau)$ is the completeness function of mass and age, shown in Figure \ref{fig:mass_age_relation}e,f, the completeness in ages $C(\tau)$ is just its projection normalized on the density in the mass-age plane:
\begin{eqnarray}
\label{eq:completeness_in_ages}
C(\tau) = \frac{\int C(M,\tau)\cdot n(M,\tau)\ \rm{d}M}{\int n(M,\tau)\ \rm{d}M} \\
 = \frac{\int n_{\rm sample}(M,\tau)\ \textrm{d}M}{\int n(M,\tau)\ \nonumber \textrm{d}M}
\end{eqnarray}
where $n(M,\tau)$ is the completeness-corrected density in the mass-age plane (Figure \ref{fig:mass_age_relation}e,f) and $n_{\rm sample}(M,\tau)$ the measured one (Figure \ref{fig:mass_age_relation}a,b).

The completeness correction does not influence significantly the average ages and the measured age spread; but increases the old age tail of the distributions.
In the \citet{siess2000} case the age distribution peaks at $\sim3$~Myr, with a broadening of 0.3~dex while with \citet{pallastahler99} the peak is at about 2~Myr with a higher broadening of 0.4~dex. However, the broader distribution of the latter in logarithmic scale vanishes in linear units, because of the younger average age. Within 1$\sigma$ the age of the population spans from 2.5 to 5~Myr for Siess, and from 1.22 to 3.22~Myr for Palla \& Stahler.
Increasing the slope of the reddening law produces a modest shift towards younger ages ($\lesssim0.1$~dex)because of the higher luminosities derived.
As mentioned in the previous section, the observed age spread does not coincide with a real star formation history, because of the unknown uncertainties in the stellar luminosities. Therefore the distributions shown in Figure \ref{fig:histogram-age} represents the combined effect of the real age spread and the apparent spread originated by other effects that scatter the measured luminosities. The latter is mainly due to stellar variability and scattered light from circumstellar material, and is probably responsible for the 7--9\% (respectively for Palla \& Stahler and Siess models) of stars with a measured age exceeding 10~Myr, percentage that increases of $\sim50\%$ when correcting for completeness. This fraction of old age sources is about 3-5 times higher than the upper limit for the residual contamination from non-members (see Section \ref{section:subsampleanalysis}).

We stress that the sources of apparent luminosity scattering we have mentioned may be responsible for the tails of the age distribution, but we do not expect the overall width of the age distribution to be significantly overestimated. In particular, our age spread agrees with the one measured in the ONC by \citet{jeffries+2007} by means of a statistical analysis of geometrically estimated radii - therefore immune of the effects of hidden binarity, variability, extinction uncertainties. This suggests that a real age spread is the main cause of the observed luminosity spread in the H-RD and that we are not significantly overestimated the width of the age distribution.

\subsection{The mass function}
The mass function (MF) of a system is defined as the number distribution of stars as a function of mass. While a general observational approach measures the so-called \emph{present day mass function} (PDMF), a cornerstone for understanding how stars form is the \emph{Initial Mass Function} (IMF), which is the mass distribution according to which stars are born. For a pre-main sequence population younger than the time required to produce a significant mass segregation, or to evolve the most massive stars into post-MS phase, the observed mass function coincides with the IMF.

The derivation of the IMF from observations always presents several sources of uncertainty, the main ones typically being unresolved companions and completeness. We can asses the completeness of our sample as a function of mass from the 2D completeness in an equivalent way as for the age completeness (eq. \ref{eq:completeness_in_ages}), projecting it on the mass scale:
\begin{eqnarray}
\label{eq:completeness_in_mass}
C(M) = \frac{\int n_{\rm sample}(M,\tau)\ \textrm{d}\tau}{\int n(M,\tau) \textrm{d}\tau}
\end{eqnarray}

\begin{figure}
\epsscale{1.1}
\plotone{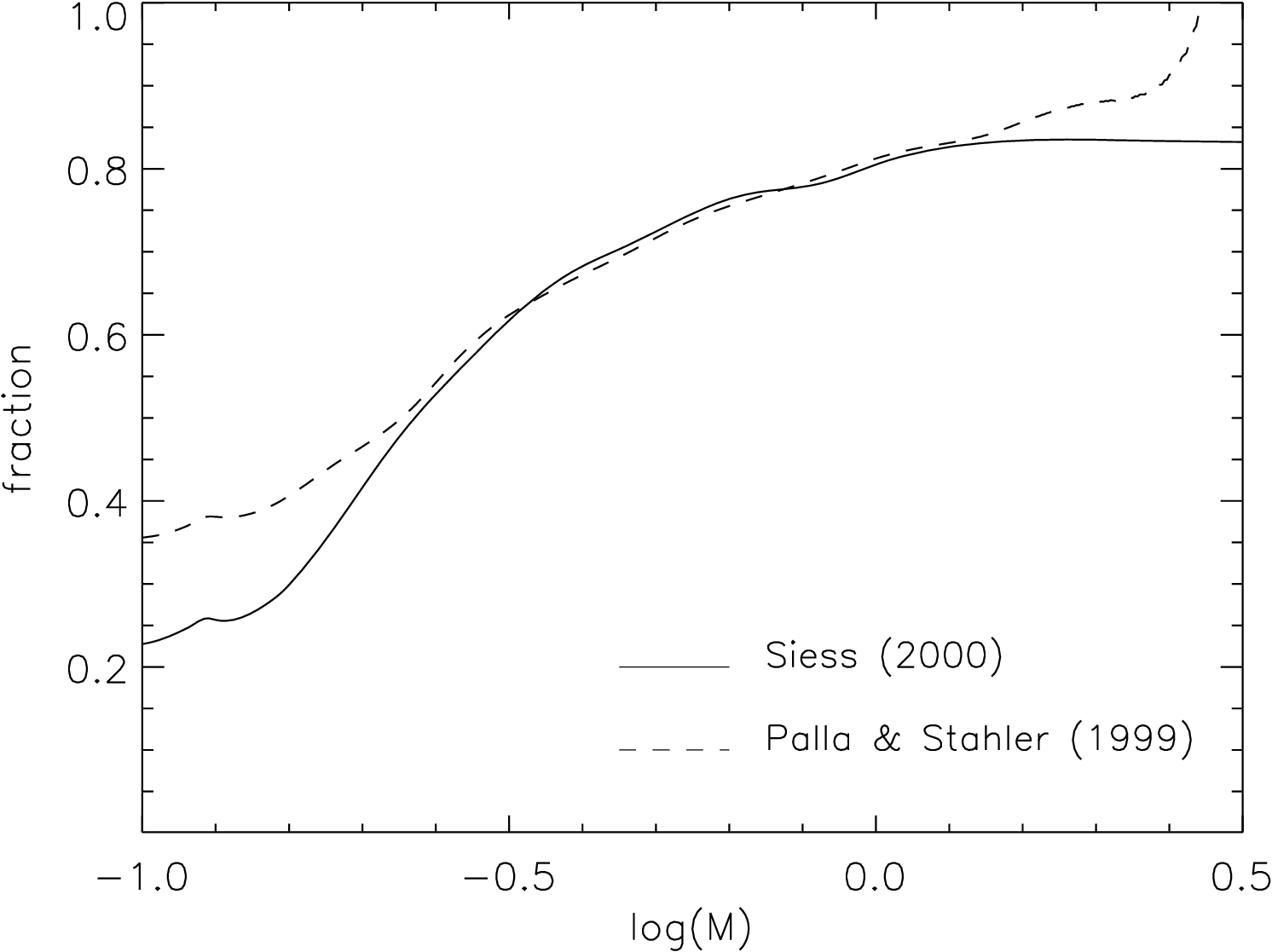}
\caption{Completeness as a function of stellar mass computed for both evolutionary models. \label{fig:completeness}}
\end{figure}

The results, computed for both evolutionary models, are presented in Figure \ref{fig:completeness}. In the intermediate-mass regime both evolutionary models provide similar completeness, whereas for lower masses the Palla \& Stahler models lead to a systematically lower completeness.
\begin{figure*}
\epsscale{1.1}
\plotone{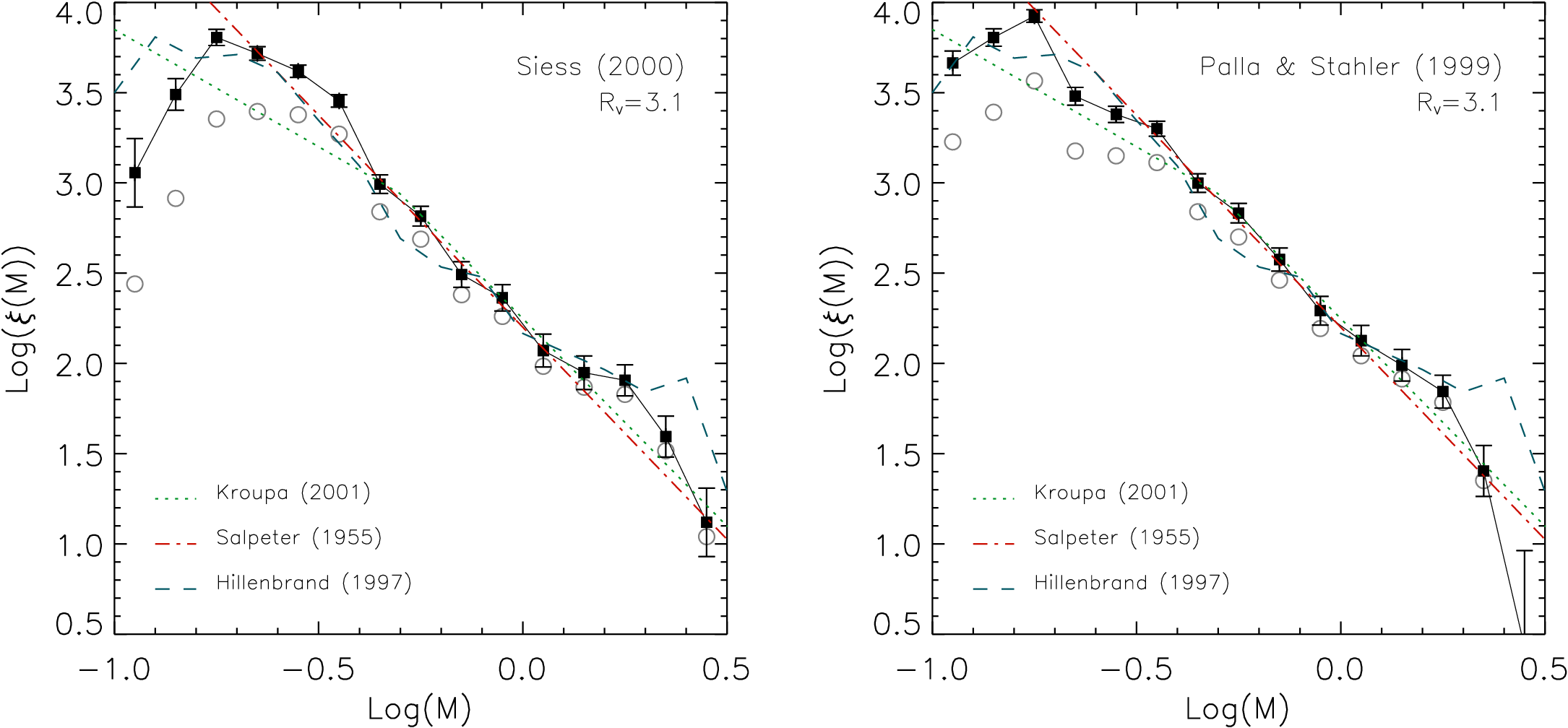}
\caption{Completeness-corrected mass functions derived for the ONC (filled squares) according to the models of \citet{siess2000} \emph{(left panel)} and \citet{pallastahler99} \emph{(right panel)}. The open circles are the same without the completeness correction. Units are number of stars per solar mass (i.e., the counts have been normalized to the width of each bin in units solar masses). The mass function for the ONC derived in H97 is superimposed to highlight the differences in the low mass range. The \citet{salpeter55} single power law IMF and the IMF of \citet{kroupa2001} are shown as well.  \label{fig:MF}}
\end{figure*}

To derive the IMF of the ONC we count the number of stars in equally spaced bins 0.1 dex wide in $\log M$ and normalize the result to the bin width in order to express the IMF in terms of $\xi(M)$ instead of $\xi(\log M)$. We correct the star counts for completeness, interpolating the function shown in Figure \ref{fig:completeness} to the mass values of the IMF.

The two panels of Figure \ref{fig:MF} show the IMF derived for  both Siess  and Palla \& Stahler\ evolutionary models. In the same figure, the H97 IMF is overplotted, showing that it follows our distribution for intermediate- and low-mass stars ($M\gtrsim0.4{\rm M_\odot}$). The \citet{kroupa2001} and
\citet{salpeter55} IMF are shown as well, evidencing the differences with our mass distributions. Specifically, Kroupa under-produces the observed  low-mass
population and does not exhibit the observed turn-over.
Salpeter over-produces the observed number of low mass stars
with the effect perhaps even larger than indicated due to the
normalization in the mid-masses rather than the high mass end.
Table \ref{table:IMF} provides the value for $\xi(M)$ and the associated uncertainty for the two cases.


\begin{table}
\caption{Initial mass function for the ONC.}
\begin{tabular}{||r|rr|rr||}
\hline
\multicolumn{1}{||c|}{ } &
\multicolumn{2}{c|}{Siess} &
\multicolumn{2}{c|}{Palla \& Stahler} \\
\multicolumn{1}{||c|}{$\log (M/{\rm M_\odot})$} &
\multicolumn{1}{c}{$\log \xi(M)$} &
\multicolumn{1}{c|}{$\sigma \log \xi(M)$} &
\multicolumn{1}{c}{$\log \xi(M)$} &
\multicolumn{1}{c||}{$\sigma \log \xi(M)$} \\ \hline
   -0.95 &     3.06 &     0.19 &     3.66 &      0.07 \\
   -0.85 &     3.49 &     0.09 &     3.81 &      0.05 \\
   -0.75 &     3.81 &     0.04 &     3.92 &      0.03 \\
   -0.65 &     3.72 &     0.04 &     3.48 &      0.05 \\
   -0.55 &     3.62 &     0.03 &     3.38 &      0.05 \\
   -0.45 &     3.45 &     0.03 &     3.30 &      0.04 \\
   -0.35 &     2.99 &     0.05 &     3.00 &      0.05 \\
   -0.25 &     2.82 &     0.05 &     2.83 &      0.05 \\
   -0.15 &     2.49 &     0.07 &     2.58 &      0.06 \\
   -0.05 &     2.36 &     0.07 &     2.29 &      0.08 \\
    0.05 &     2.07 &     0.09 &     2.13 &      0.08 \\
    0.15 &     1.95 &     0.09 &     1.99 &      0.09 \\
    0.25 &     1.91 &     0.09 &     1.84 &      0.09 \\
    0.35 &     1.59 &     0.11 &     1.40 &      0.14 \\
    0.45 &     1.12 &     0.19 &     0.43 &      0.53 \\ \hline
\end{tabular}
\label{table:IMF}
\end{table}

As we mentioned in Section \ref{section:age_distribution}, the luminosity spread we measure can be slightly overestimated due to physical and observational biases such as stellar variability, hidden binarity, scattered light, and uncertainties in the positioning of the members in the H-RD. However, since the evolutionary tracks in the H-RD are nearly vertical in the mass range we investigate, and given that, as mentioned, a true age spread is probably the dominant source of luminosity spread, we do not expect that our IMF is significantly affected by these biases.

The two sets of models produce significant differences in the IMF for        low-mass stars,  $\log (M)\lesssim0.3{\rm M_\odot}$. For both models the IMF shows a flattening at low masses. This feature is real, given our accurate assessment of completeness. However,  while for Palla \& Stahler the change of slope from the general power-law is relatively modest and in agreement with the  Kroupa mass distribution,  the Siess model leads to a clear turn-over  below $0.2{\rm M_\odot}$, compensated by an apparent overabundance of stars in the range $0.2{\rm M_\odot}<M<0.3{\rm M_\odot}$.
Similar differences in the shape of the IMF\ below $0.2$${\rm M_\odot}$, due to differences between theoretical models, temperature scales and bolometric corrections, were also reported by \cite{hillenbrand2000}.

We tested also the changes when $R_V$ is increased (not shown in the Figure), finding that this does not introduce significant modifications in the shape of the IMF. This is not surprising, considering that the reddening law affects the luminosity of stars, shifting the point in the H-R\ diagram along the $y-$axis in proportion to the color excesses $E(V-I)$, whereas for most of the stars in this temperature and luminosity range the evolutionary tracks  are nearly vertical.

With respect to the IMF\ presented by \cite{Muench+02}, derived from a Monte Carlo best-fit to the observed $K-$band luminosity function, we confirm the power-law increase with slope similar to the $x=-1.35$ value of \cite{salpeter55}, for masses larger than $M\simeq0.6~$${\rm M_\odot}$. However, the break they found at $M\simeq0.6~$${\rm M_\odot}$ occurs in our case at lower masses, $M\simeq0.3~$${\rm M_\odot}$ or less. This is consistent with the fact that there ONC members are known to present $K-$band excess due to circumstellar material, which biases the fluxes towards brighter values which are erroneously interpreted as higher masses than the underlying stars truly possess.

Thanks to the highest number of sources present in the subsample used to derive the IMF, as well as our completeness correction, we are able to constrain the position of the IMF peak at low stellar masses with a higher confidence than previous works. As discussed in \citet{briceno2008}, the presence of a peak in the stellar distribution at spectral types $\sim$M3 (corresponding roughly with masses $M\sim0.3$M$_\odot$) is measured for all the young populations of the Orion complex, suggesting a common origin for all these populations. On the contrary, other star forming regions such as Taurus and IC348 present IMFs increasing down to the hydrogen burning limit, and in the case of Taurus, showing evidences of stellar overabundance at $M\sim0.6$--$0.8$M$_\odot$. According to our results, we confirm the presence of a peak in the very-low mass regime, but this is located lower masses, closer to 0.2${\rm M_\odot}$.

In conclusion, our analysis shows  that especially at the lowest stellar masses, and presumably also in the entire brown dwarf regime, the actual shape of the IMF\ depends strongly on the model assumptions. Our current understanding of the pre-main-sequence stellar evolution suffers from the lack of a general consensus on the best model to adopt. By combining accurate and simultaneous multicolor photometry with spectroscopy it is possible to derive critical information  to constrain the models, as we have tried to show in this paper. An extension  of this analysis in the substellar mass range will be enabled by the study of the HST ACS photometry (\emph{Robberto et al., in preparation})  obtained for the HST Treasury Program on the Orion Nebula Cluster, to which this ground-based observational effort  also belongs.

\section{Summary}
\label{section:conclusion}
In this work we perform an analysis of the simultaneous multi-band optical photometry from WFI observations presented in \citet{paperI},  deriving a new H-R diagram of the ONC\ population in the low- and intermediate-mass range and the corresponding mass function.

From the comparison of the observed color-color diagrams obtained from $B$,$V$,$I$ and 6200\AA\ narrow-band filters, with results of synthetic photometry, we find evidences that the intrinsic colors for the ONC differ from those of main-sequence dwarfs in the M-type regime, but they are shifted towards bluer colors, in a similar way as predicted by atmosphere models assuming $\log g$ values from a PMS isochrone.
Nevertheless, the atmosphere models we use are still unable to match accurately the photospheric emission of the ONC; we therefore calibrate empirically the intrinsic color scale based on our ONC photometry, limiting to the known members with modest extinction and no accretion detected from H$\alpha$ photometry.

The color-color diagrams of the population also suggests that the typical Galactic value of the reddening parameter  $R_V=3.1$ is more compatible with our data of the ONC than a higher value of $R_V=5.5$.

Our derivation of intrinsic colors, in particular, allows us to improve the accuracy of the derived $A_V$ for the ONC members, a critical step for a precise positioning of the stars in the HRD. We find evidence of $B$-band excess for a fraction of sources in the color-color diagrams, which we attribute to stellar accretion. We develop a method that, using 3 bands simultaneously, allows us to disentangle the effects on the broad-band colors of reddening and accretion luminosity. We test the accuracy of this method finding that while it leads to a robust and more unbiased determination of $A_V$, it is still insufficient to provide a strong constrain on the accretion luminosity.

The derivation of all the stellar parameters still needs an independent estimate of \teff. We use the spectral catalog of \citet{hillenbrand97}, as well as new spectral types that we present here, and a sample M-type stars classified by means of narrow-band photometry (Paper~I). This allows us to have at disposal the most complete sample of stars with known spectral type in the ONC. We study how the choice of the spectral type-\teff\ relation is critical to remove systematic trends in $A_V$ as a function of \teff. We find the the \citet{luhman2000} temperature scale improves over the \citet{cohencuhi79} scale.

We derive the bolometric luminosities for $\sim1000$ members and position them the HRD, down to the hydrogen-burning limit. The revised lower distance we used of 414~pc compared to previous works leads to systematically lower luminosities; this is compensated for in M-type stars, where we predict higher $L$, caused by higher measured extinctions for late-type stars. This is due to our improvements in the intrinsic colors and our method to account also for accretion effects in the observed colors.

We compute the 2D completeness function of our sample of members in the HRD, including both photometric detection and the spectral subsample.
Using the evolutionary models of \citet{siess2000} and \citet{pallastahler99} we assign ages and masses to all the members.

We extract the age distribution from the different choice of  models. In the case of Siess the average ages turn out to be higher ($\sim3$~Myr) than Palla \& Stahler ($\sim2$~Myr) with a slightly lower spread (0.3~dex vs 0.4~dex).  We find a mass-age correlation if one uses the Palla \& Stahler models, with  high-mass stars appearing older than low-mass stars. This trend, however, disappears when a careful statistical analysis of the completeness of our stellar sample in the mass-age plane is performed.

We derive the IMF of our sample, corrected for detection incompleteness down to the hydrogen burning limit; the IMF turns out to be model-dependant in the very-low mass range, with Siess predicting fewer low-mass stars and a turn-over at $M\simeq 0.2{\rm M_\odot}$ while using Palla \& Stahler a more modest flattening is present.

\

\acknowledgements

N.D.R. kindly acknowledges financial support from the German Aerospace Center (DLR) through grant 50~OR~0401.

K.G.S. acknowledges support from NSF Career grant AST-0349075 and a Cottrell Scholar award from the Research Corporation.

This work was made possible in part by GO\ program 10246 of the {\it Hubble Space Telescope}, which is operated by the Space Telescope Science Institute.

This work was made possible through the Summer Student Program of the Space Telescope Science Institute.

{\it Facilities:} \facility{ESO, HST} 

\end{document}